\documentclass{pasj00}
\draft

\begin{document}
\SetRunningHead{Kaneda et al.}{Dust and PAHs in NGC~4125}
\Received{//}
\Accepted{//}

\title{Properties of dust and PAHs in the hot plasma of the elliptical galaxy NGC~4125 revealed with AKARI and Spitzer}

\author{%
   Hidehiro \textsc{Kaneda},\altaffilmark{1}
   Daisuke \textsc{Ishihara},\altaffilmark{1}
   Takashi \textsc{Onaka},\altaffilmark{2}
   Toyoaki \textsc{Suzuki},\altaffilmark{3}
   Tatsuya \textsc{Mori},\altaffilmark{1}\\  
   Shinki \textsc{Oyabu},\altaffilmark{1}
   and
   Mitsuyoshi \textsc{Yamagishi},\altaffilmark{1}}
 \altaffiltext{1}{Graduate School of Science, Nagoya University \\
Nagoya, Aichi 464-8602}
 \email{kaneda@u.phys.nagoya-u.ac.jp}
 \altaffiltext{2}{Department of Astronomy, Graduate School of Science, University of Tokyo, \\
Bunkyo-ku, Tokyo 113-0003}
 \altaffiltext{3}{Institute of Space and Astronautical Science, \\
Japan Aerospace Exploration Agency, Sagamihara, Kanagawa 252-5210}



\maketitle

\begin{abstract}

We present the spatial distributions of dust and polycyclic aromatic hydrocarbons (PAHs) in the elliptical galaxy NGC~4125, revealed by AKARI and Spitzer. NGC~4125 is relatively bright in the dust and the PAH emision for elliptical galaxies, although it certainly possesses diffuse interstellar hot plasma indicated by the high spatial resolution X-ray data of Chandra. We investigate how the dust and PAHs interact with the X-ray plasma or avoid the interaction by comparing their spatial distributions. We find that the distributions of the PAHs and dust are different from each other, both showing a significant deviation from a smooth stellar distribution. The PAH emission predominantly comes from a dust lane, a compact dense molecular gas region in the galactic center, where the PAHs are likely to have been protected from the interaction with the X-ray plasma. The dust emission has more extended structures similar to the distribution of the X-ray plasma, suggesting their interaction to some extent. We also discuss a possible origin of the dust and PAHs in the galaxy. 
\end{abstract}

\section{Introduction}
Many elliptical galaxies possess an appreciable amount of X-ray-emitting hot plasma providing a harsh interstellar environment for the survival of polycyclic aromatic hydrocarbons (PAHs) and dust grains. Dust with a size of 0.1 $\mu$m is destroyed through sputtering on such a short timescale as $10^6$ yr for the plasma density and temperature of $10^{-2}$ cm$^{-3}$ and $10^7$ K typical of elliptical galaxies (Draine \& Salpeter 1979; Tielens et al. 1994). Recently, Micelotta et al. (2010) showed that the destruction timescale of typical PAHs (50 C atoms, size$\sim$ 6\AA) in hot plasma with temperature higher than 3$\times10^4$ K is so short as $10^2$ yr for the above plasma parameters, which is $10^{2-3}$ times shorter than lifetimes for dust with the same size. Thus the presence of dust grains in X-ray elliptical galaxies would imply that they were supplied very recently and the presence of PAHs would even suggest that there are some ways for them to avoid interaction with the hot plasma.

Despite such a hostile environment, it has been found that a significant fraction of X-ray-emitting elliptical galaxies contain a considerable amount of dust, which cannot be explained solely from replenishment by old stars (e.g. Knapp et al. 1989; Goudfrooij \& de Jong 1995; Temi et al. 2004; Temi et al. 2007). Surprizingly some of them even show the presence of PAHs (e.g. Kaneda et al. 2005, 2008a). Any clear signature of interaction of the dust and PAHs with X-ray plasma has not yet been obtained either from the relations of their emission luminosities among galaxies or from their spatial distributions within a galaxy. The former probably implies that the dust and PAHs are supplied by transient events rather than continuous events, their amounts depending on how recently the events happened. The latter is mainly due to a paucity of spatial information on the dust and PAH emission. For NGC~4589, the deep mid-IR spectral mapping with the Spitzer/IRS has revealed the distinctive spatial distribution of the PAH 11.3 $\mu$m emission perfectly following the minor-axis dust lane (Kaneda et al. 2010). The distributions of the far-IR dust continuum and the PAH 17 $\mu$m emission are extended more widely in directions similar to each other and entirely different from the stellar distribution. Although the galaxy is detected in the X-ray, it is so faint that the distribution of diffuse hot plasma to be compared with those of the dust and PAHs cannot be derived.  

In this paper, we report on the mid-IR spectral mapping and near- to far-IR imaging observations of NGC~4125, an X-ray elliptical galaxy categorized as E6pec (Filho et al. 2002). The mid-IR spectral mapping data are obtained with the Infrared Spectrograph (IRS; Houck et al. 2004) onboard Spitzer (Werner et al. 2004), while the imaging data are obtained with the Infrared Camera (IRC: Onaka et al. 2007) and the Far-Infrared Surveyor (FIS; Kawada et al. 2007) onboard AKARI (Murakami et al. 2007). NGC~4125 has an intermediate X-ray luminosity (Log $L_X$ = 40.94 erg s$^{-1}$) among the ROSAT galaxies cataloged by O'Sullivan et al. (2001). The galaxy habors an emission line nucleus with a low-ionization nuclear emission-line region (LINER) spectrum (Heckman, Balick, \& Crane 1980; Willner et al. 1985). Schweizer \& Seitzer (1992) concluded that NGC~4125 is a $(6-8)\times10^9$ yr old remnant of a major disk-disk merger based on the $UBV$ colors and the amount of merger-induced fine structures. Indeed the galaxy shows ionized gas emission along the major axis, which does not follow simple circular motions (Bertola et al. 1984). The kinematic profiles of stars suggest that there is a colder substructure in the innermost center (Pu et al. 2010). Verdoes Kleijn \& de Zeeuw (2005) found a dust lane of size $\timeform{1''.3}$ in the center of the galaxy. Contrary to the suggested link between merging and X-ray brightness, the galaxy is not so X-ray-bright as hot gas-rich elliptical galaxies (Fabbiano \& Schweizer 1995). Yet NGC~4125 certainly possesses diffuse interstellar hot plasma as indicated by the high spatial resolution X-ray data of Chandra and XMM-Newton (Humphrey et al. 2006; Fukazawa et al. 2006) and an appreciable amount of far-infrared (IR) dust (Knapp et al. 1989; Goudfrooij \& de Jong 1995; Temi et al. 2004; Temi et al. 2007), thus is an ideal elliptical galaxy to search for signature of interaction of the dust and PAHs with the X-ray plasma.

The contents of the paper are based on our new datasets of the IRS spectral maps and the AKARI images. NGC~4125 was also observed within the framework of the Spitzer SINGS legacy program (Kennicutt et al. 2003), although details of the individual galaxy have not been reported in a paper. As for the IRS observations, we performed spectral mapping observations, which were $2-8$ times deeper and $2-3$ times wider than the SINGS observations, depending on the spectrometer modules. Our AKARI image data provide new information at the photometric bands that the IRAC and MIPS do not have, especially at wavelengths of 11 $\mu$m, 15 $\mu$m, and 90 $\mu$m. Below we adopt 22.2 Mpc for the distance to NGC~4125 (Tonry et al. 2001; Jensen et al. 2003), where 1 kpc corresponds to the angular size of $\timeform{9.3''}$.


\section{Observations}
The Spitzer/IRS spectral mapping observations were performed in May 2009, while the AKARI/IRC and FIS imaging observations were carried out in November 2006 and May 2007. The observation log is listed in table 1.  The Spitzer data were taken in part of our Guest Observers (GO5) program (PI: H. K.; program ID: 50369). Figure 1a shows the slit positions of the Short-Low (SL; 5.2--14.5 $\mu$m) and Long-Low (LL; 14--36 $\mu$m) modules. For $\sim 1'\times 1'$ area around the galactic center, we made a $33\times 2$ grid for SL with a ramp duration of 60 s for each exposure, and a $13\times 2$ grid for LL with a ramp duration of 30 s for each. We shifted the slit in directions perpendicular and parallel to its length by $\timeform{1''.85}$ and $\timeform{3''.0}$, respectively, for SL, and by $\timeform{5''.1}$ and $\timeform{9''.0}$, respectively, for LL. We repeated the same spectral mapping observations two times for LL. Starting with the basic calibrated data (pipeline version S18.7.0) reduced by the Spitzer Science Center, we extracted spectral image data by the CUBISM software (version 1.7; Smith et al. 2007a) following standard procedures. From the spectral image data, we created the spectra of the central $15''\times15''$ region. Figure 1b shows the resultant spectra, where the background spectra created from eight (SL) and four (LL) slit-aperture data at both ends of each mapping area are subtracted. We observed a small mismatch between the continuum levels of the SL and LL spectra and hence applied a scale factor of 1.2 to the LL data to obtain their agreement at the overlapping wavelengths.  

\begin{table*}
\caption{Observation Log}
\begin{center}
\begin{tabular}{lcccc}
\hline\hline
Instrument&Date&Observing mode&Observation ID&Details\\ \hline
AKARI/IRC&Nov 12, 2006&IRC02&1400041&{\it N3}, {\it N4}, {\it S7}, {\it S11} band imaging\\
AKARI/IRC&May 11, 2007&IRC02&1402057&{\it L15}, {\it L24} band imaging\\
AKARI/FIS&Nov 13, 2006&FIS01&1400044&Scan speed: $8''$ s$^{-1}$, Reset interval: 2.0 s\\
Spitzer/IRS&May 13, 2009&IrsMap&26091264&see text\\
\hline
\end{tabular}
\end{center}
\end{table*}

\begin{figure}
\FigureFile(60mm,60mm){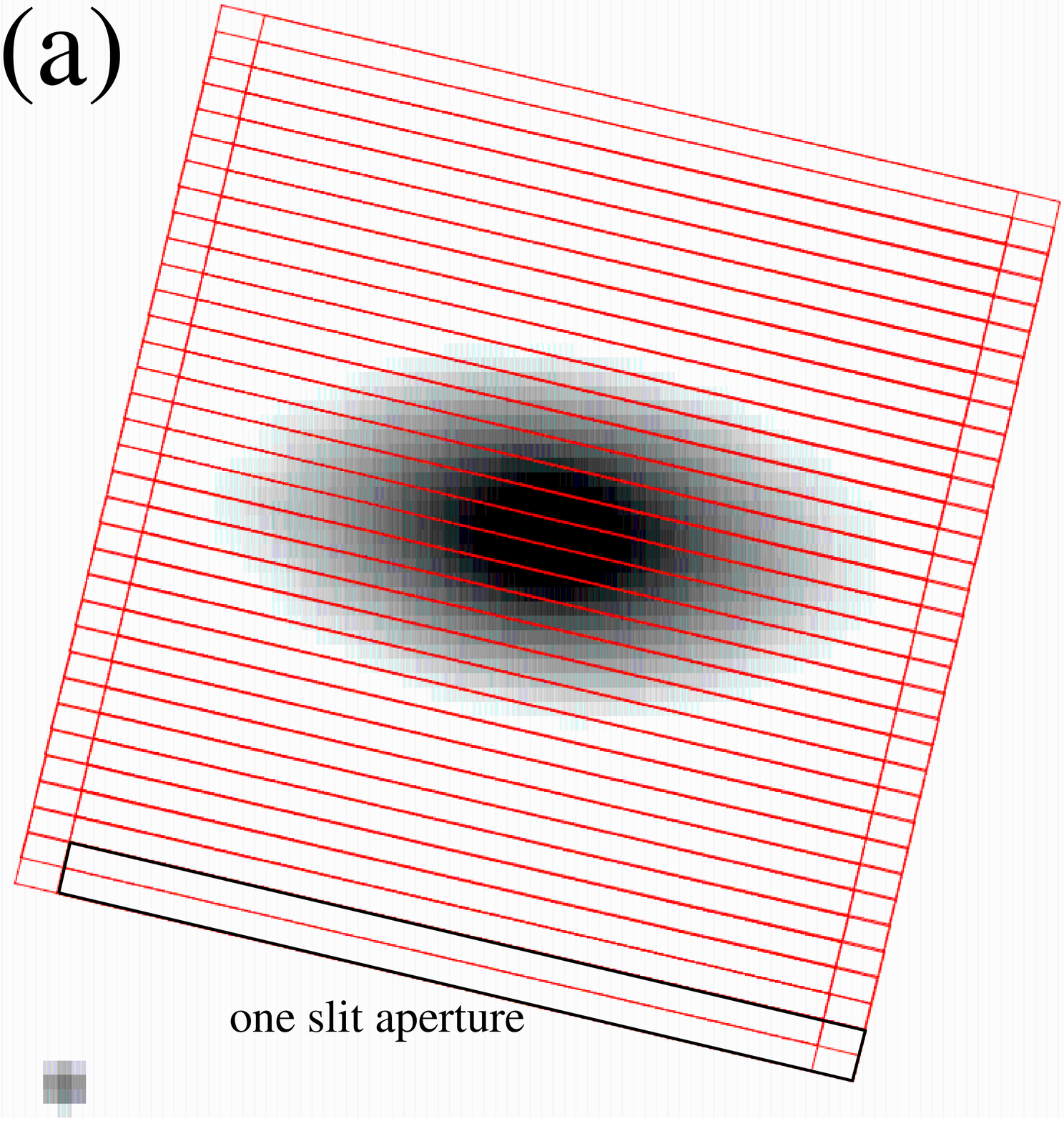}\FigureFile(60mm,60mm){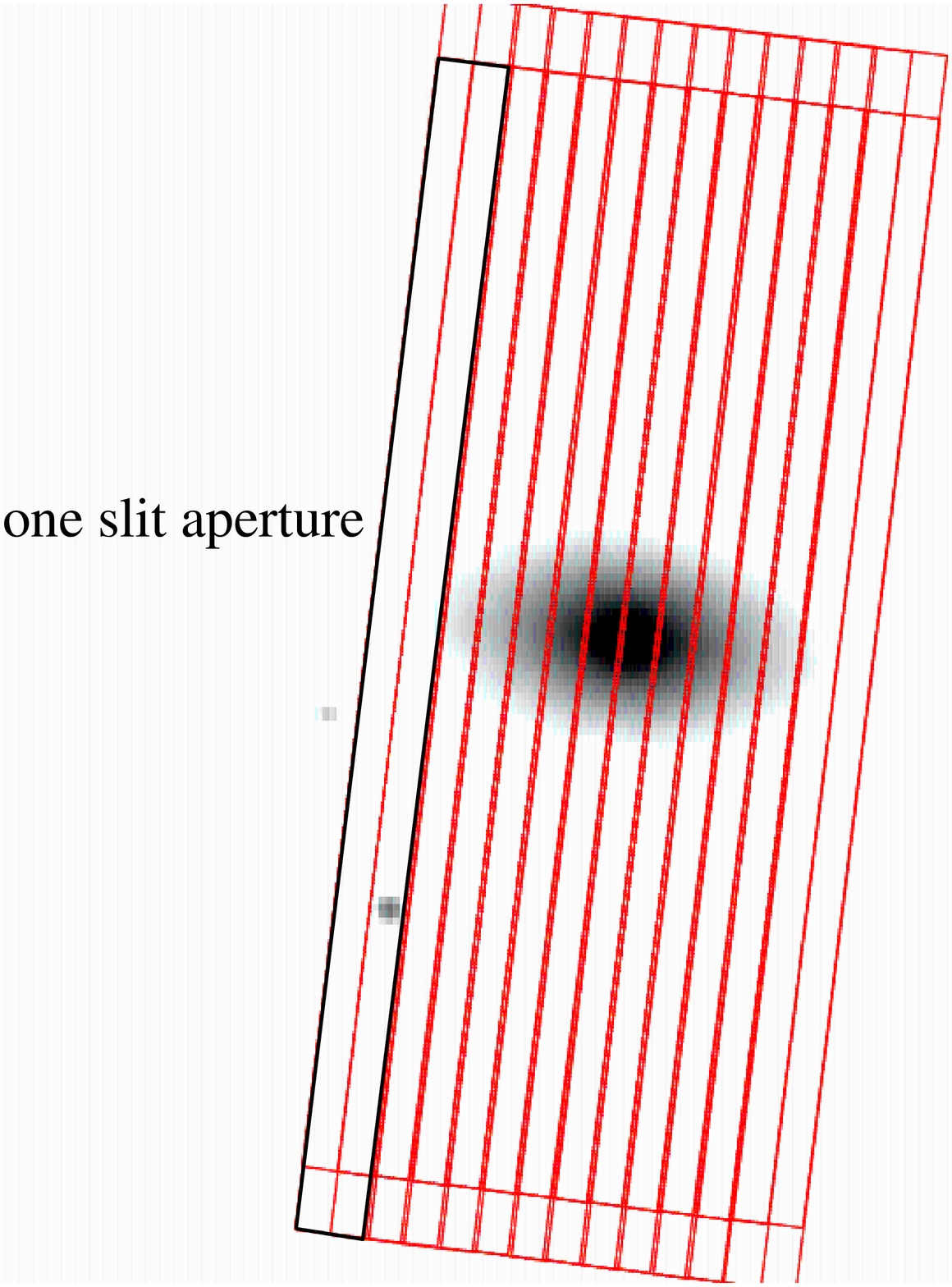}\\
\FigureFile(120mm,120mm){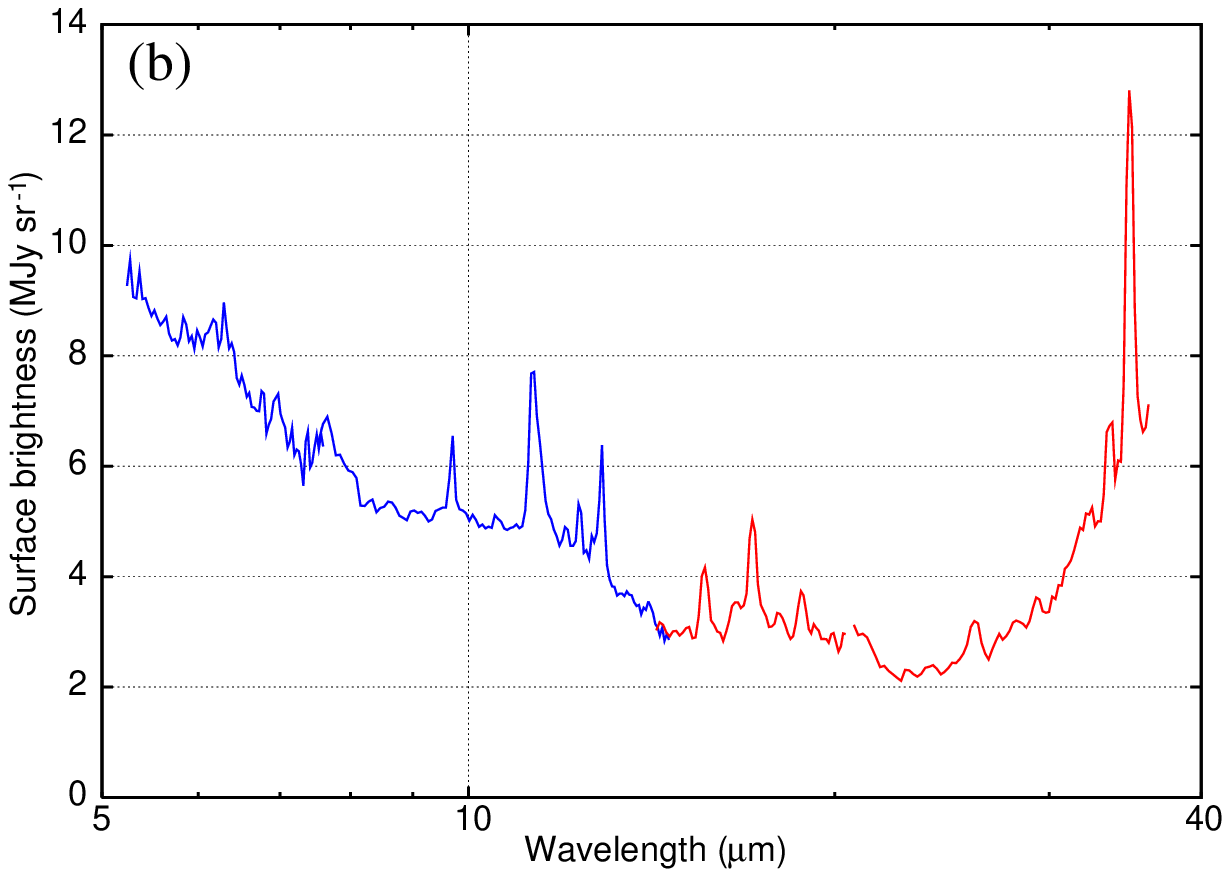}\\
\caption{(a) illustrative views of IRS SL (left) and LL (right) slit positions for the spectral mapping observations of NGC~4125, superposed on the 2MASS J-band image of NGC~4125. The image sizes are $1'.3\times 1'.3$ for SL and $3'\times 3'$ for LL. The north is up and the east is to the left. (b) The background-subtracted IRS SL (blue) and LL (red) spectra of the central $15''\times15''$ region of NGC~4125.}
\end{figure}

The AKARI data were taken in part of the AKARI mission program ``ISM in our Galaxy and Nearby Galaxies'' (ISMGN; PI: H. K.). With the IRC, we obtained the {\it N3} (reference wavelength of 3.2 $\mu$m), {\it N4} (4.1 $\mu$m), {\it S7} (7.0 $\mu$m), {\it S11} (11.0 $\mu$m), {\it L15} (15.0 $\mu$m), and {\it L24} (24.0 $\mu$m) band images using a standard staring observation mode. The images were created by using the IRC imaging pipeline software version 20071017 (see IRC Data User Manual; Lorente et al. 2007 for details). The background levels were estimated by averaging values from multiple apertures placed around the galaxy, and subtracted from the images. With the FIS, we obtained {\it N60} (centered at a wavelength of 65 $\mu$m), {\it WIDE-S} (90 $\mu$m), {\it WIDE-L} (140 $\mu$m), and {\it N160} (160 $\mu$m) band images using a standard slow-scan observation mode, a 2-round-trip slow scan with a scan speed of $8''$ s$^{-1}$. A region approximately $10'\times 10'$ around the galaxy was covered with a scan map. The images were processed from the FIS Time Series Data using the AKARI official pipeline developed by the AKARI data reduction team (Verdugo et al. 2007) with custom procedures to remove cosmic-ray-induced detector artifacts. The background levels were estimated from data taken near the beginning and the end of the slow scan observation and subtracted from the images.

\section{Results}
\subsection{Spitzer mid-IR spectral maps}
As seen in figure 1b, we clearly detect the prominent PAH 11.3 $\mu$m feature and [SiII] 34.8 $\mu$m emission line, together with several emission lines such as H$_2$S(3) at 9.7 $\mu$m, [NeII] at 12.8 $\mu$m, [NeIII] at 15.6 $\mu$m, H$_2$S(1) at 17.0 $\mu$m, [SIII] at 18.7 and 33.5 $\mu$m, the PAH 17 $\mu$m broad feature, and the mid-IR dust continuum emission at wavelengths longer than 27 $\mu$m. Figure 2 shows the contour maps of the spectral components for the central $45''\times45''$ area, where the bin sizes of spectral images are set to be $\timeform{1''.85}$ for SL and $\timeform{5''.08}$ for LL. We applied smoothing with a boxcar kernel of 3 pixels in width ($\sim 5''$ for SL and $\sim 15''$ for LL) for every spectral map. In table 2, we list the wavelength ranges used for extracting each spectral component. Note that the [NeII] and H$_2$S(1) line emissions are blended with the underlying PAH 12.7 $\mu$m and 17 $\mu$m features, respectively. For the PAH 6.3 and 7.7 $\mu$m features, we could not obtain reliable spectral maps because the slopes of the underlying continua are steep and the features are broad. 

\begin{figure}
\FigureFile(50mm,50mm){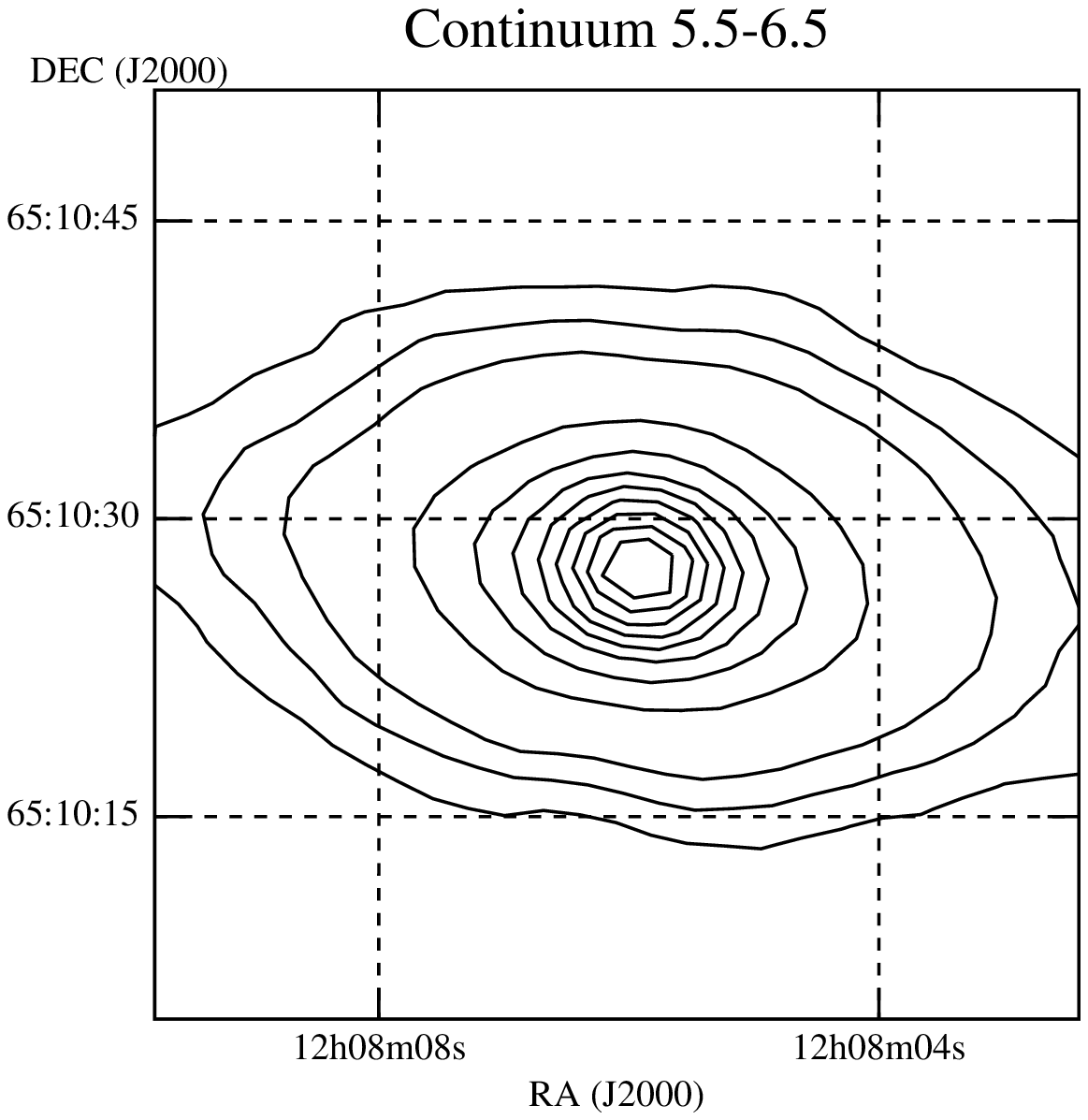}\FigureFile(50mm,50mm){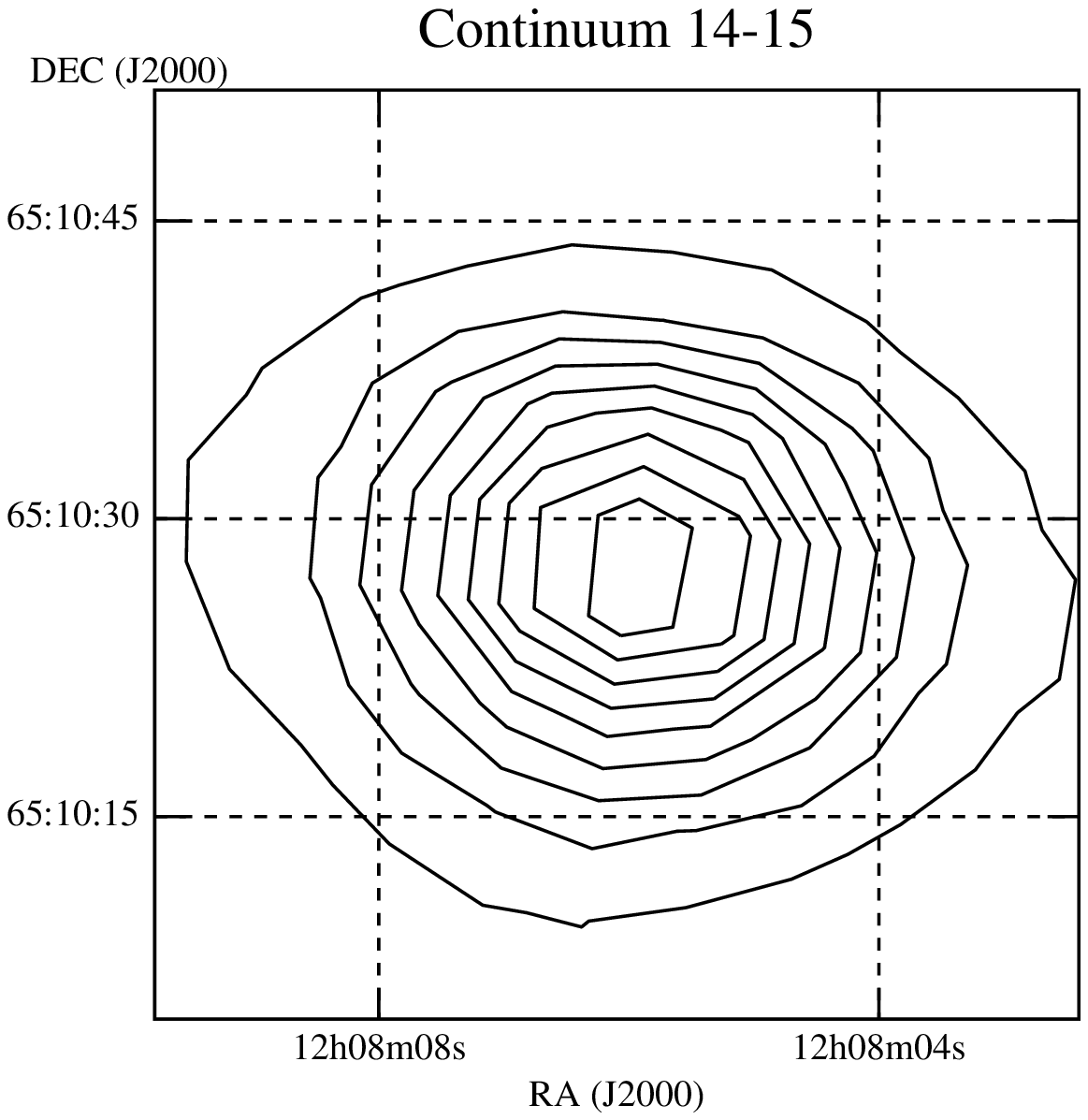}\FigureFile(50mm,50mm){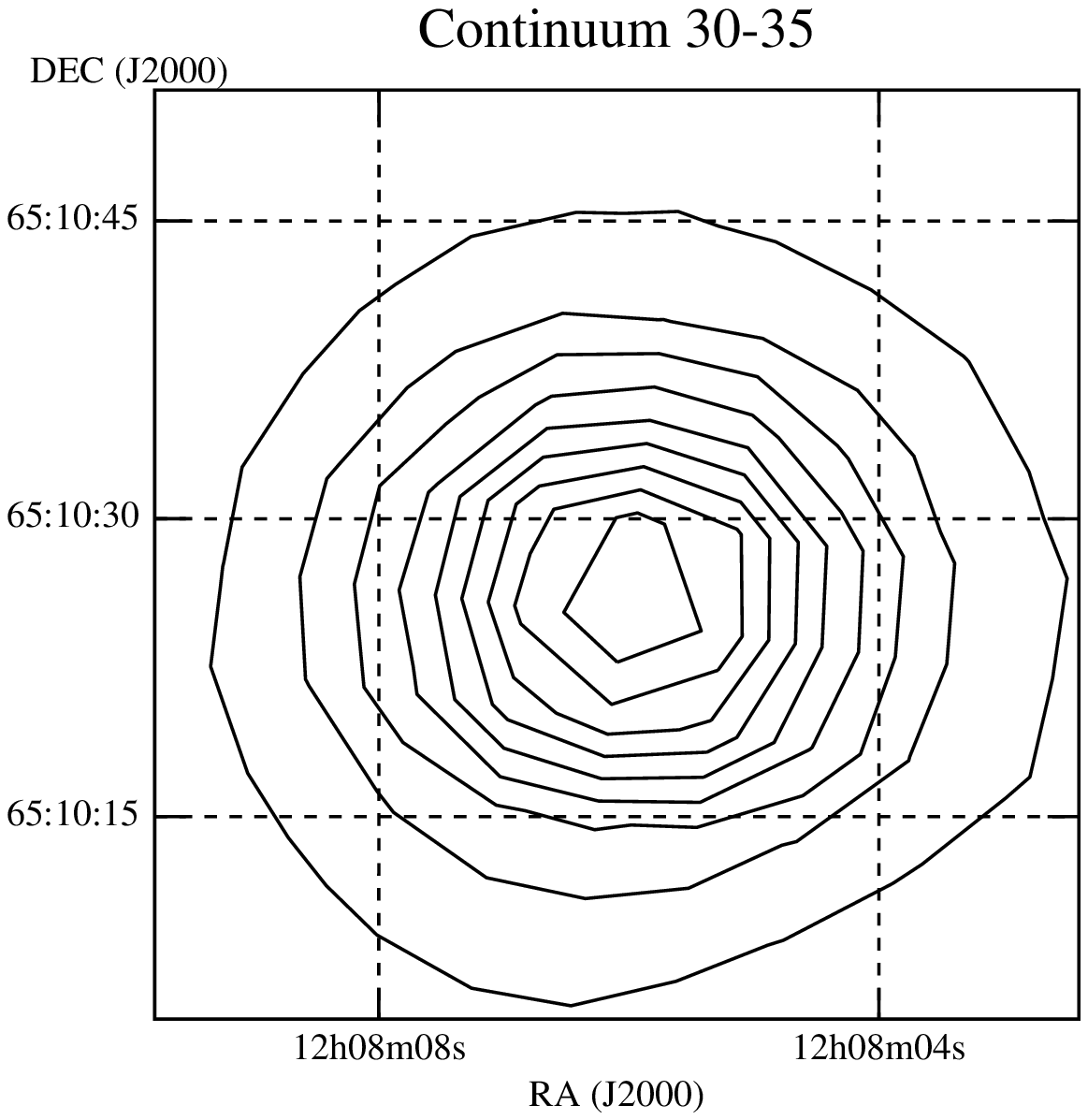}\\
\FigureFile(50mm,50mm){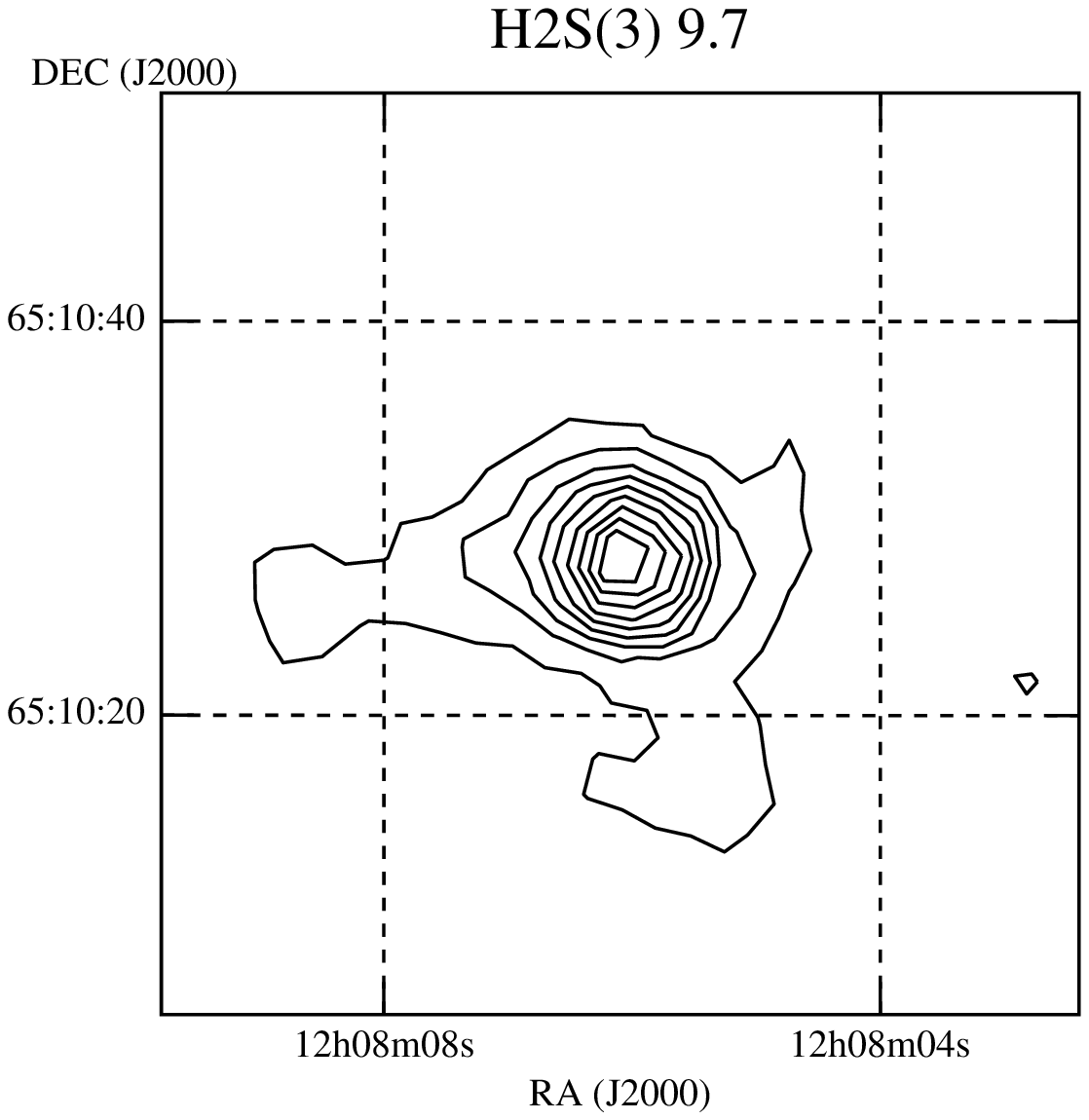}\FigureFile(50mm,50mm){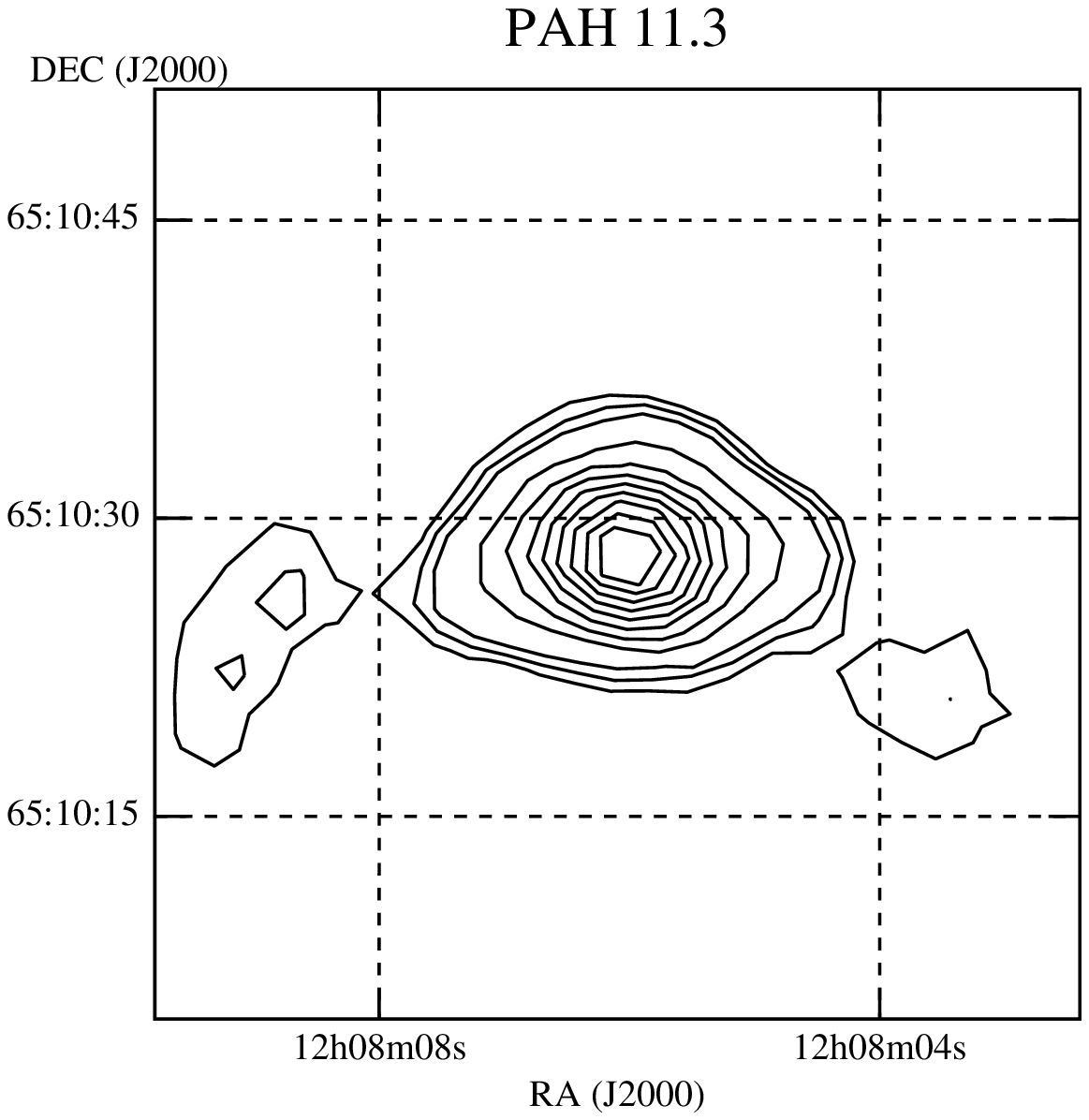}\FigureFile(50mm,50mm){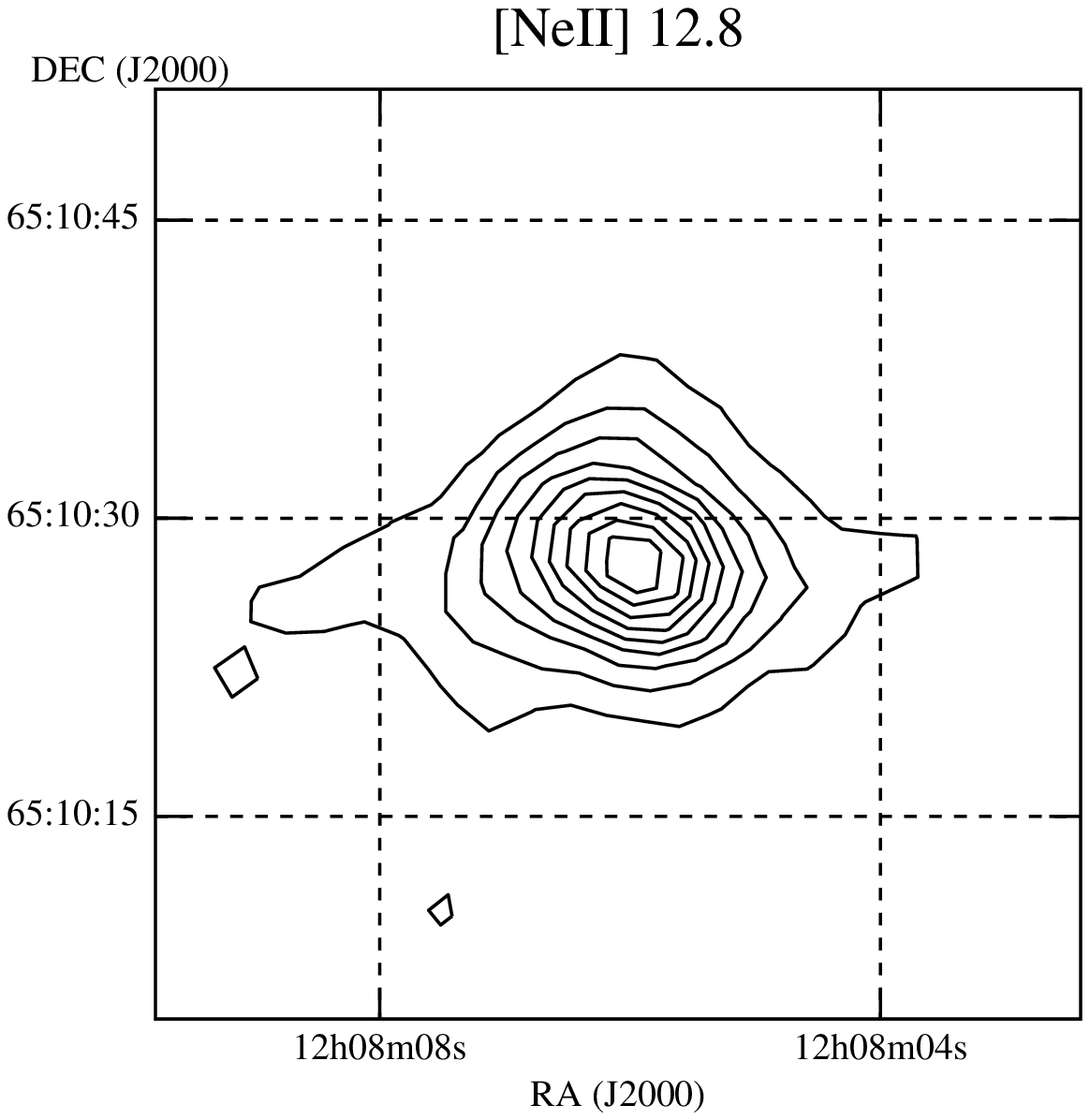}\\
\FigureFile(50mm,50mm){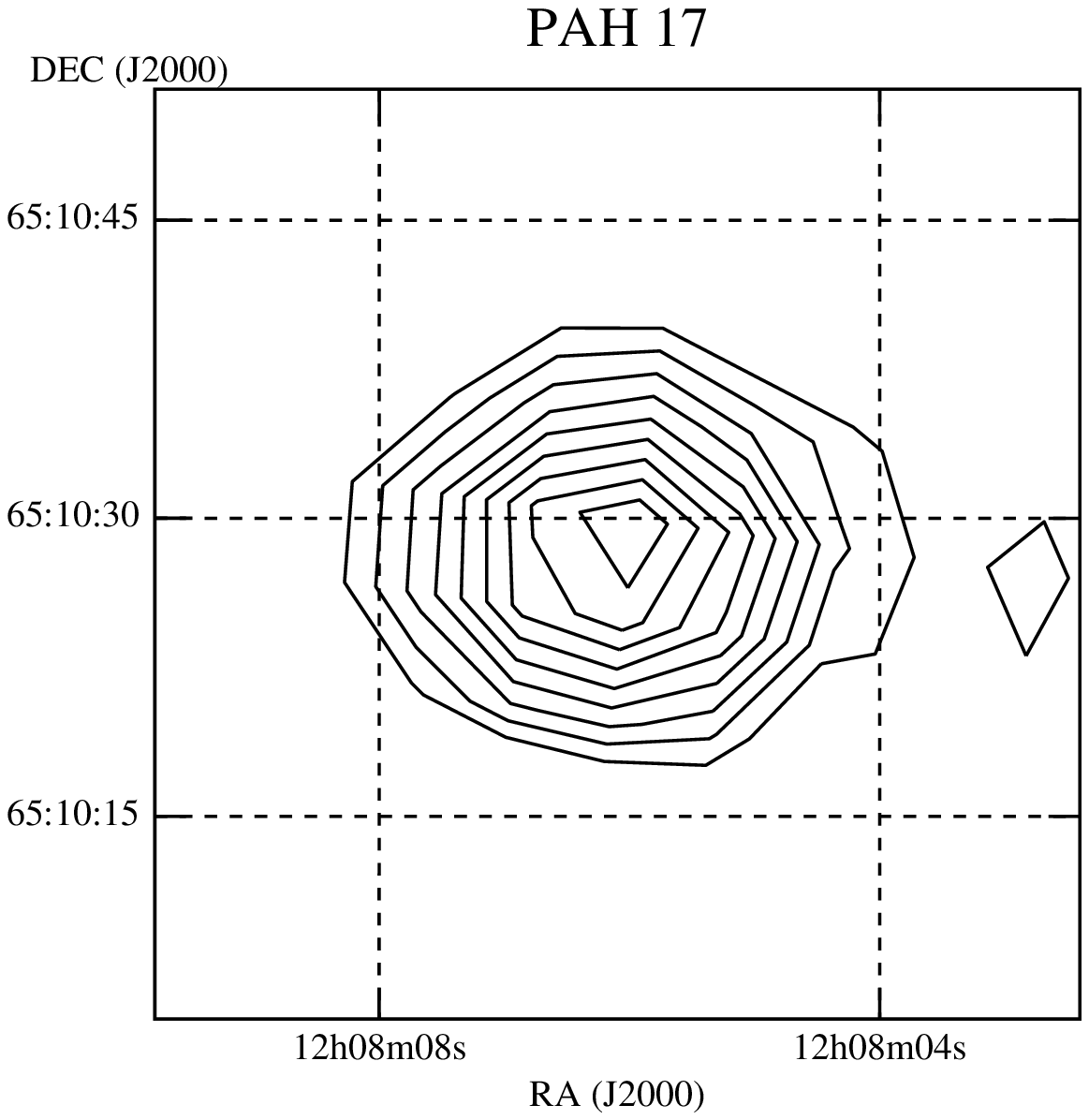}\FigureFile(50mm,50mm){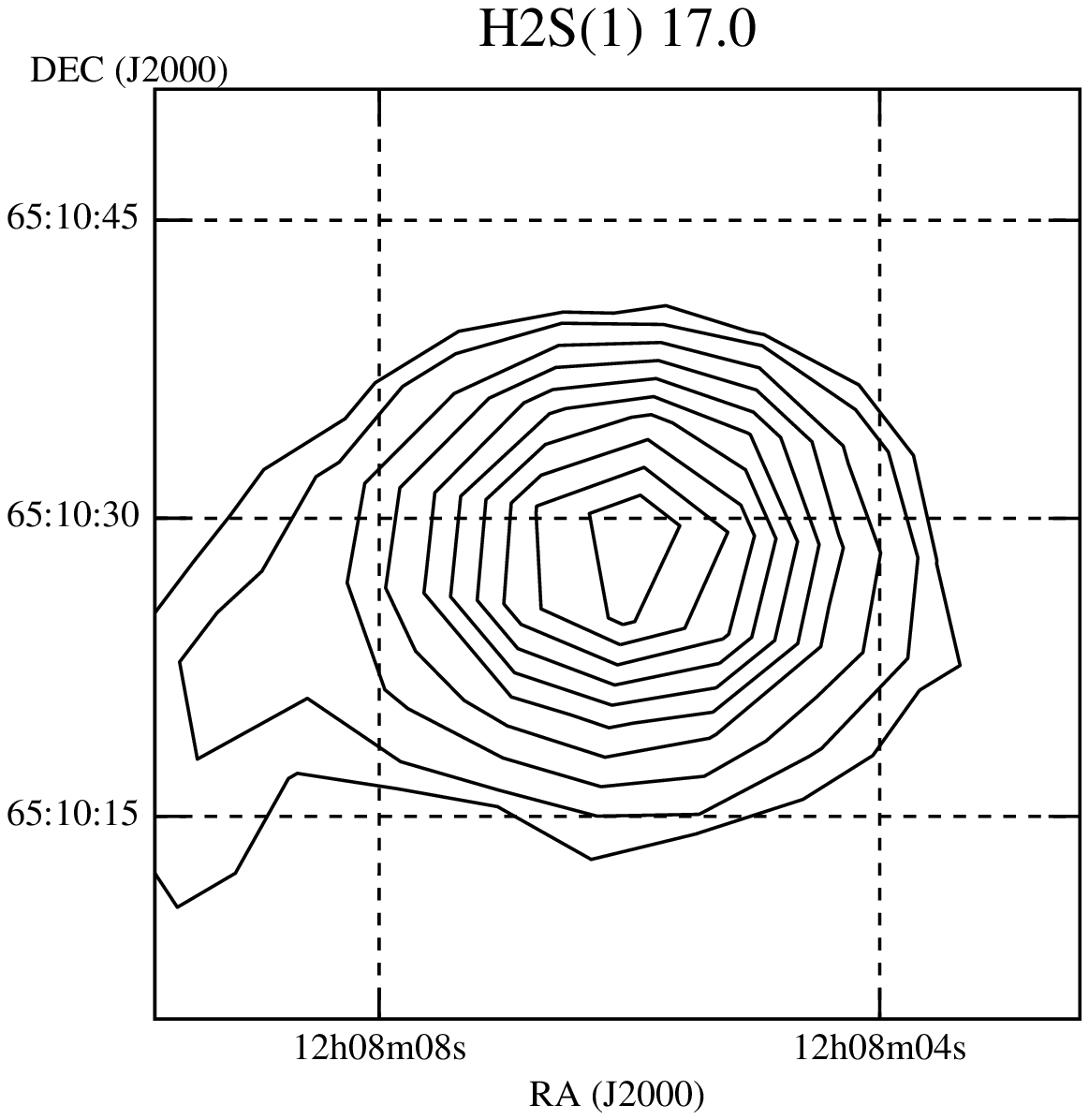}\FigureFile(50mm,50mm){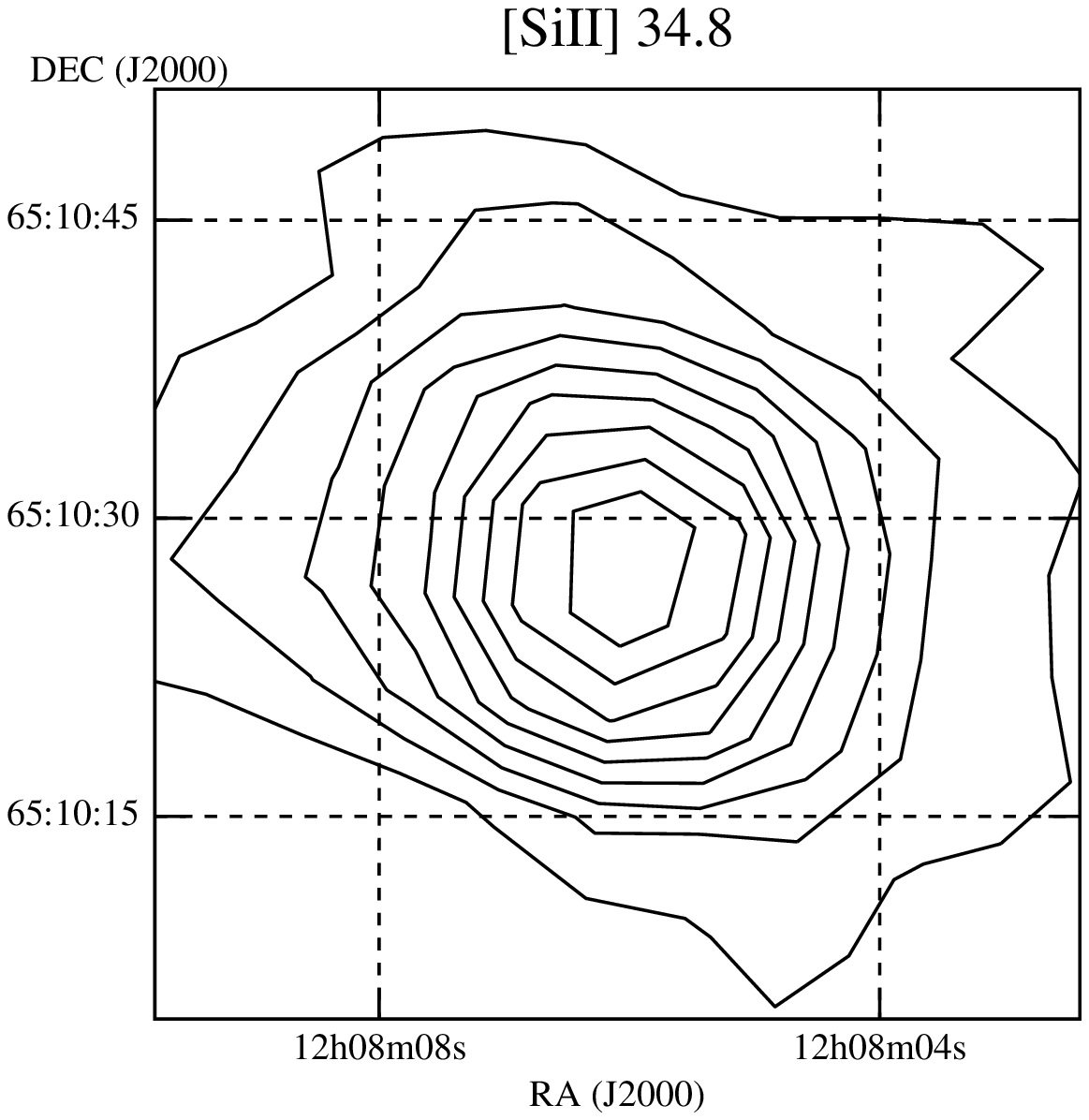}\\
\caption{Spiter/IRS spectral maps of the $5.5-6.5$ $\mu$m continuum, $14-15$ $\mu$m continuum, $30-35$ $\mu$m continuum, H$_2$S(3) 9.7 $\mu$m, PAH 11.3 $\mu$m, [NeII] 12.8 $\mu$m, PAH 17 $\mu$m, H$_2$S(1) 17.0 $\mu$m, and [SiII] 34.8 $\mu$m emissions. For each map, the contours are drawn on a linear scale from 90 to 10 \% of the peak surface brightness. The 5 \% and 7 \% level contours are added for the $5.5-6.5$ $\mu$m continuum and PAH 11.3 $\mu$m map, while the 5 \% level contour is added for the H$_2$S(1) map. }
\end{figure}

\begin{table*}
\caption{Wavelength ranges used for deriving the spectral maps in Fig.2}
\begin{center}
\begin{tabular}{lrr}
\hline\hline
Component & Signal & Background\\
 & ($\mu$m) & ($\mu$m)\\ \hline
$5.5-6.5$ $\mu$m continuum&$5.5-6.5$&\dots  \\
$14-15$ $\mu$m continuum&$14.4-14.9$&\dots  \\
$30-35$ $\mu$m continuum&$30.0-34.5$&\dots  \\
H$_2$S(3) 9.7 $\mu$m&$9.6-9.8$&$9.0-9.5$, $9.9-10.9$ \\
PAH 11.3 $\mu$m&$11.0-11.8$&$9.9-11.0$, $11.8-12.3$  \\
$[$NeII$]$ 12.8 $\mu$m&$12.8-12.9$&$13.0-13.2$\\
PAH 17 $\mu$m&$16.5-16.9$, $17.3-18.6$& $14.4-15.5$, $19.5-20.7$\\
H$_2$S(1) 17.0 $\mu$m&$16.9-17.3$&$14.4-15.4$, $15.8-16.3$, $19.4-20.8$\\
$[$SiII$]$ 34.8 $\mu$m&$34.5-35.5$&$33.8-34.5$, $35.5-36.2$ \\
\hline
\end{tabular}
\end{center}
\end{table*}

As seen in figure 2, the distributions of the continuum emissions at wavelengths of 5.5--6.5 $\mu$m and 14--15 $\mu$m show an elliptical shape, which is consistent with the stellar distribution in the 2MASS image (figure 1a), while that at 30--35 $\mu$m is of a more circular shape. These distributions support that the continuum emission at wavelengths shorter than 20 $\mu$m is dominated by photospheric emission from old stars as also suggested by the spectral slope, and that the mid-IR dust continuum emission at wavelengths longer than 27 $\mu$m is distributed differently from the stellar emission. In fact it is generally known that near- to mid-IR emission from an elliptical galaxy is dominated by photospheric emission from low-mass giant stars with temperatures of 3000 K to 5000 K (Athey et al. 2002; Xilouris et al. 2004; Temi et al. 2008).  

The distributions of the PAH 11.3 $\mu$m and 17 $\mu$m as well as those of H$_2$S(3) and S(1) are relatively compact as compared to the stellar distributions at 5.5--6.5 $\mu$m and 14--15 $\mu$m. The PAHs and H$_2$ are extended in a similar direction with the position angle (P.A.) of $\sim$90$^{\circ}$, where P.A. is measured from the north through the east. The atomic gas line emissions of [NeII] (at relatively high brightness levels) and [SiII] are extended similarly in the direction of P.A. of $\sim$70$^{\circ}$, but differently from the PAHs and H$_2$. More interestingly, they are extended in different directions from the major axis P.A. $82.5^{\circ}$ of the stellar distribution (Jarrett et al. 2003). Quantitative comparisons of spatial distributions among all the relevant images are summarized later in the discussion (section 4.3).

In order to clearly visualize the difference in size of the distribution between each spectral component, spectra are compared in figure 3, which are created by integrating the areas of $5''\times5''$, $10''\times10''$, $15''\times15''$, and $25''\times25''$. As seen in figure 3a, the intensities of some spectral features with respect to the underlying continuum intensities change from the inner to the outer area. Figure 3b shows the flux ratios of the innermost ($5''\times5''$) to the outermost ($25''\times25''$) spectrum. Since the smallest area is comparable to the beam size of the IRS/LL module, we cannot discuss the change of the continuum shape. But the change of the PAH features and the line emissions relative to the underlying continua (i.e. their equivalent widths) can be discussed from figure 3b. It is clear that only the PAH 11.3 $\mu$m and H$_2$S(3) emissions, and possibly the H$_2$S(1) emission change significantly, while all the atomic gas line emissions show no significant changes, implying that the PAHs and H$_2$ are the most centrally concentrated. In particular, the difference between the PAH 11.3 $\mu$m and [NeII] 12.8 $\mu$m emission is conspicuous; there is no change in the [NeII] line equivalent width from the inner to the outer area, demonstrating that the distribution of the [NeII] emission is extended more than the PAH 11.3 $\mu$m emission. The structures around 14 $\mu$m and 19 $\mu$m are probably not real because they are located near the edges of the 1st order of the SL module and the 2nd order of the LL module, respectively.   

\begin{figure}
\FigureFile(150mm,100mm){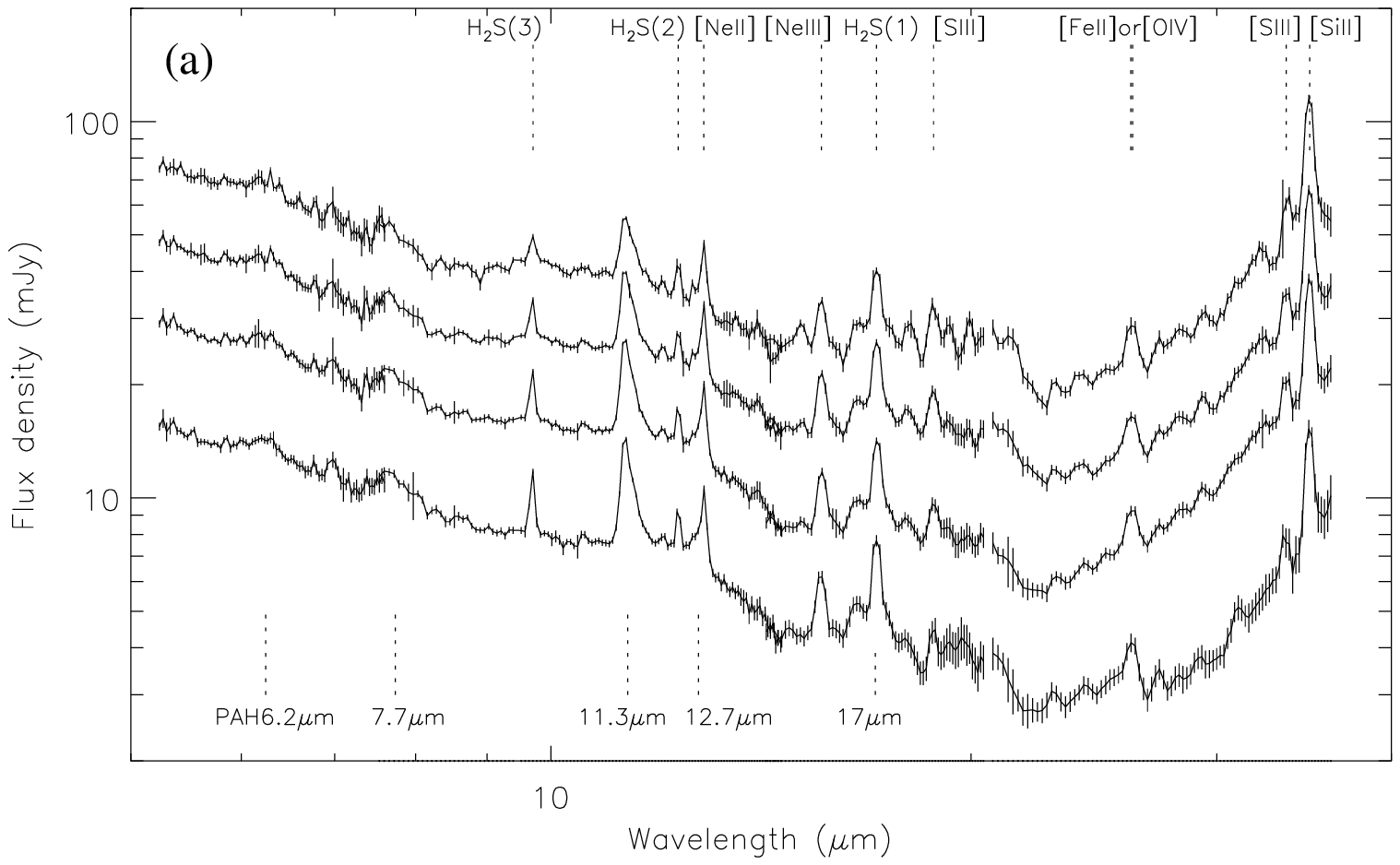}\\
\FigureFile(150mm,100mm){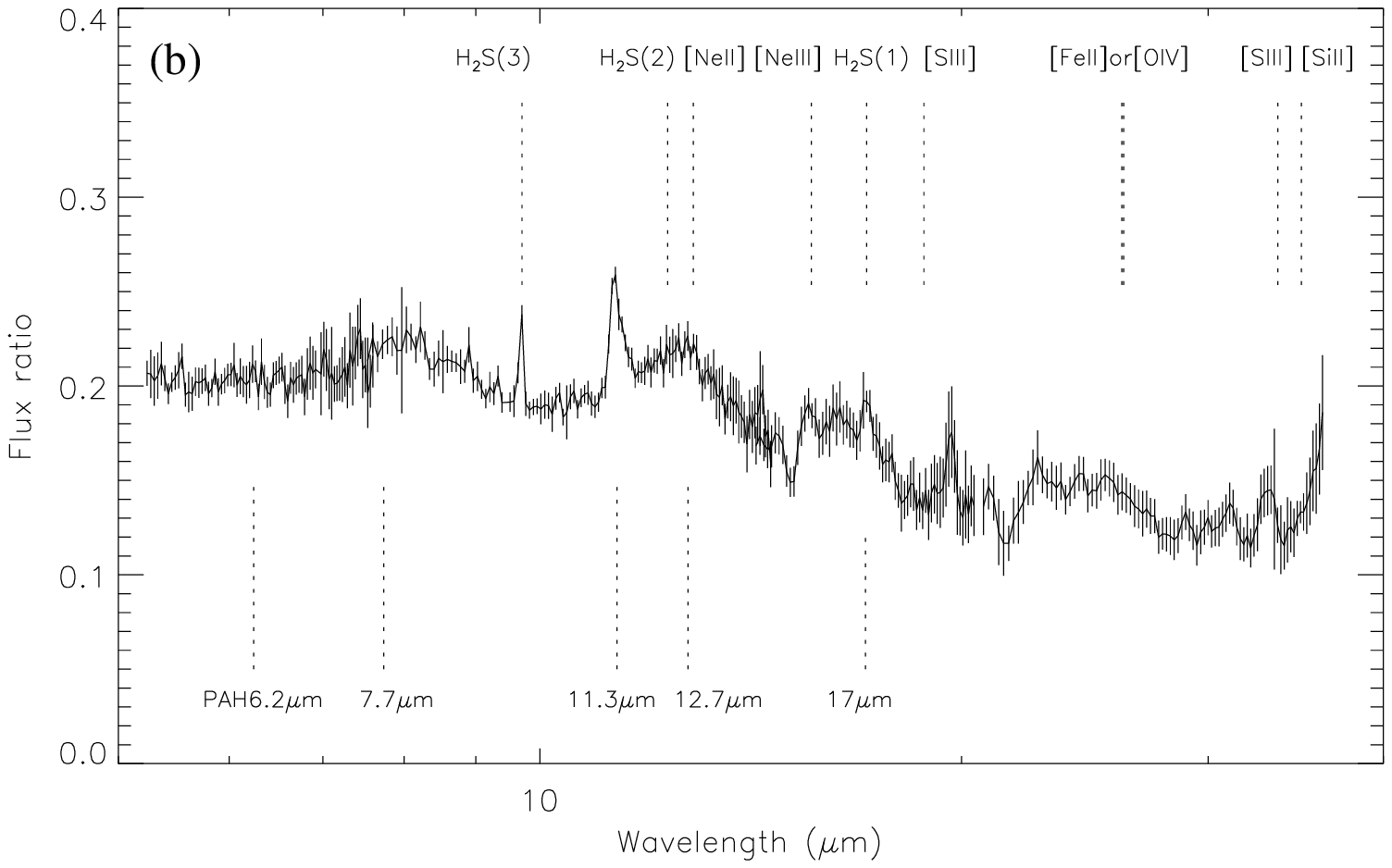}\\
\caption{(a) Spitzer/IRS spectra created by integrating data within the central $5''\times5''$, $10''\times10''$, $15''\times15''$, and $25''\times25''$ areas in the ascending order. (b) The ratio of the spectrum of $5''\times5''$ to that of $25''\times25''$.}
\end{figure}

\subsection{AKARI near- to far-IR images}
The near- to far-IR 9-band images of NGC~4125 are shown in figure 4. We do not use the {\it N160} data in the following discussion because we could not obtain significant detection. For every map in figure 4, the image size is set to be the same ($\sim 3'\times 3'$). The bin sizes of the images are $\timeform{1''.46}$ for {\it N3} and {\it N4}, $\timeform{2''.34}$ for {\it S7} and {\it S11}, $\timeform{2''.5}$ for {\it L15} and {\it L24}, $\timeform{7''.5}$ for {\it N60} and {\it WIDE-S}, and $\timeform{12''.5}$ for {\it WIDE-L}. As seen in figure 4, the spatial distribution changes considerably with the photometric band. The {\it N3}, {\it N4} and {\it S7} images show smooth distributions tracing the stellar component of the galaxy. However, the {\it S11} and {\it L15} bands start to deviate from the smooth stellar distribution. The {\it L24} band image shows a relatively faint compact emission. These changes of the images in the mid-IR are quite reasonable judging from the above Spitzer spectra. As for the 11 $\mu$m-bright spot located at $\sim1'$ to the southwest from the center, there is no counterpart found in the SIMBAD database. This can also be recognized in the 15 $\mu$m and 24 $\mu$m bands and hence is probably a real mid-IR source. However the 11 $\mu$m emission bridging the gap between the galaxy and the source is probably caused by a low-level ghost signal of the galaxy emission due to the internal reflection within the beam splitter, which are known to appear at about $1'$ from an original signal. We ignore this southwest emission structure below.

\begin{figure}
\FigureFile(50mm,50mm){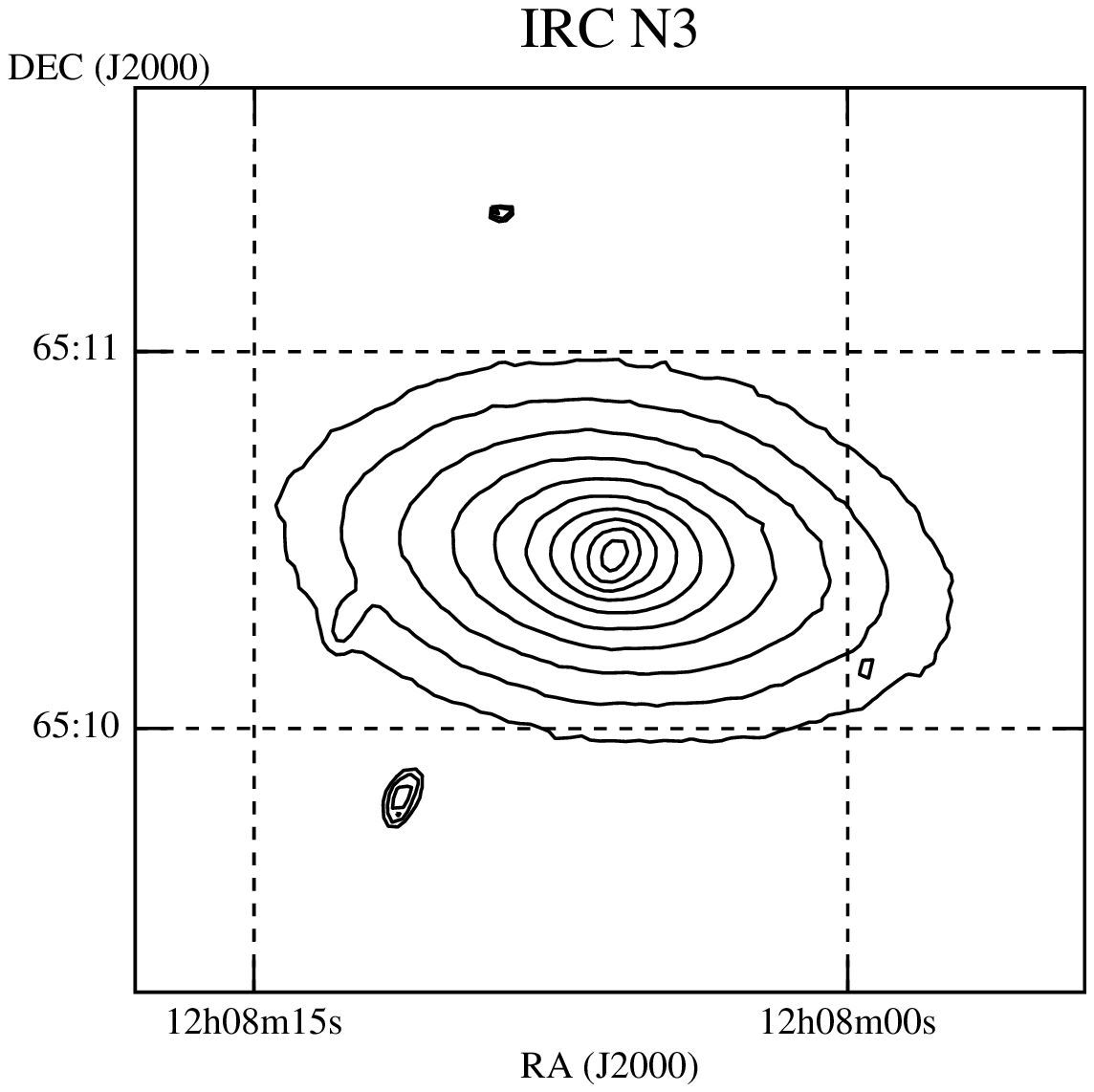}\FigureFile(50mm,50mm){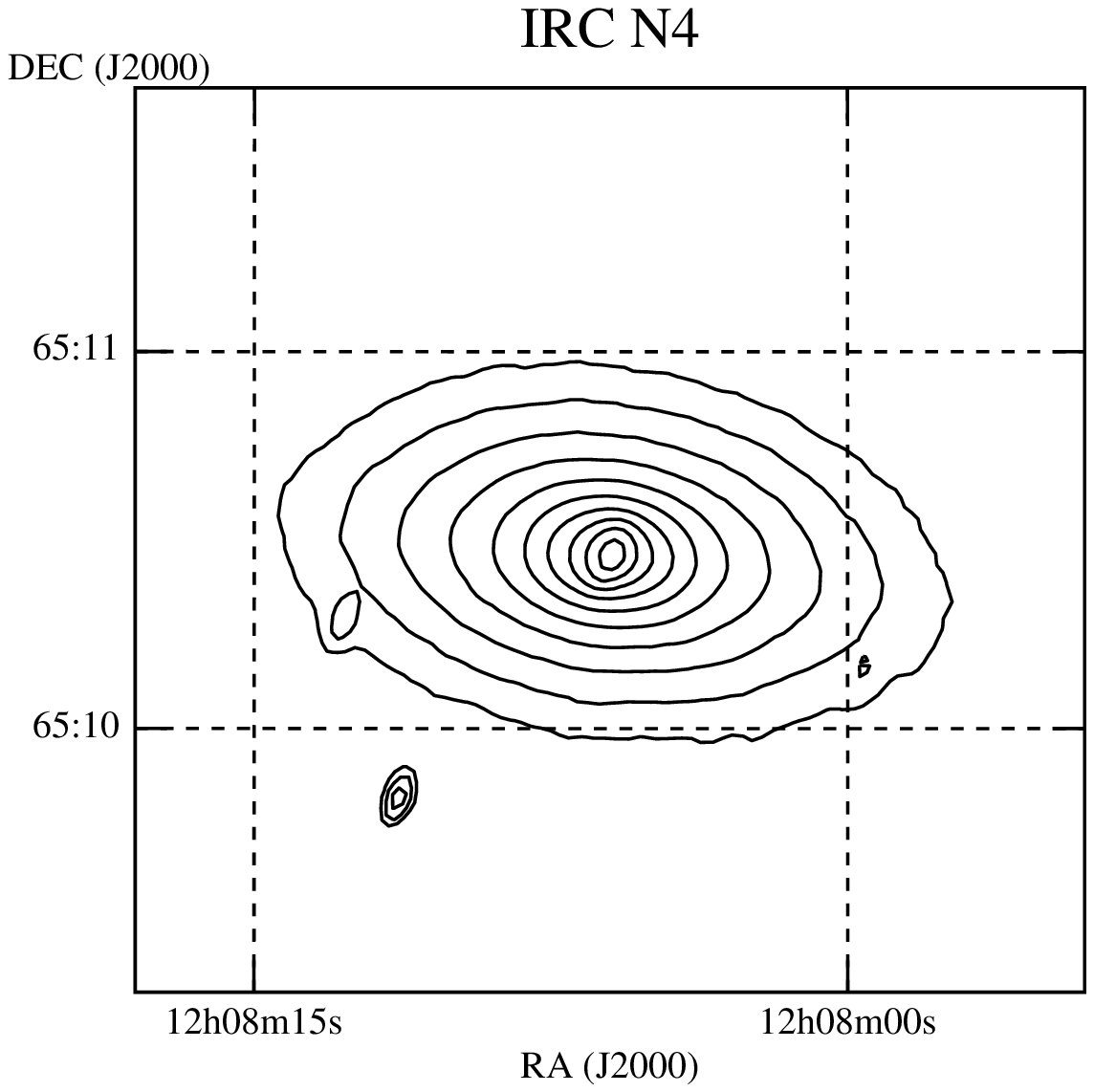}\FigureFile(50mm,50mm){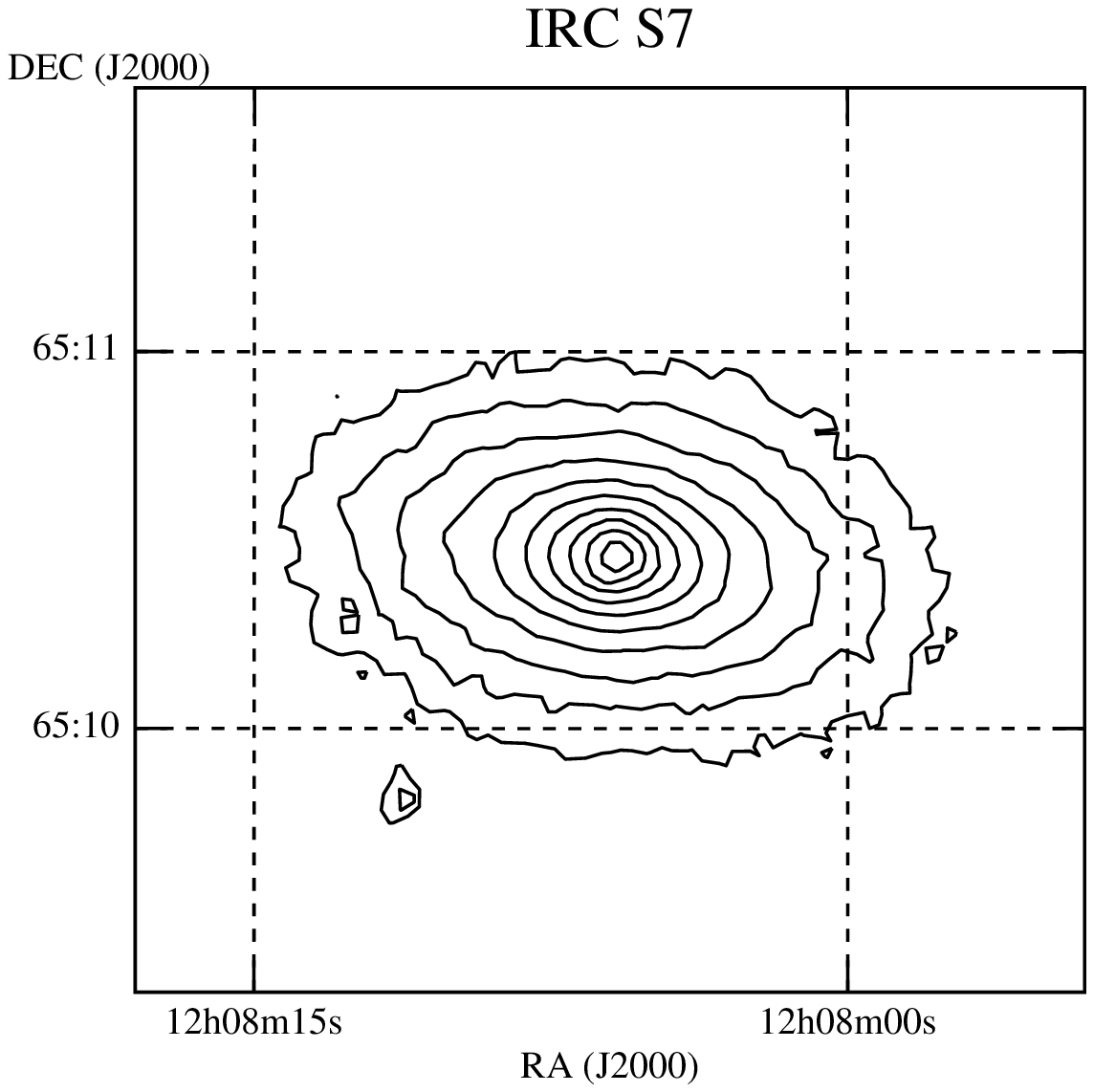}\\
\FigureFile(50mm,50mm){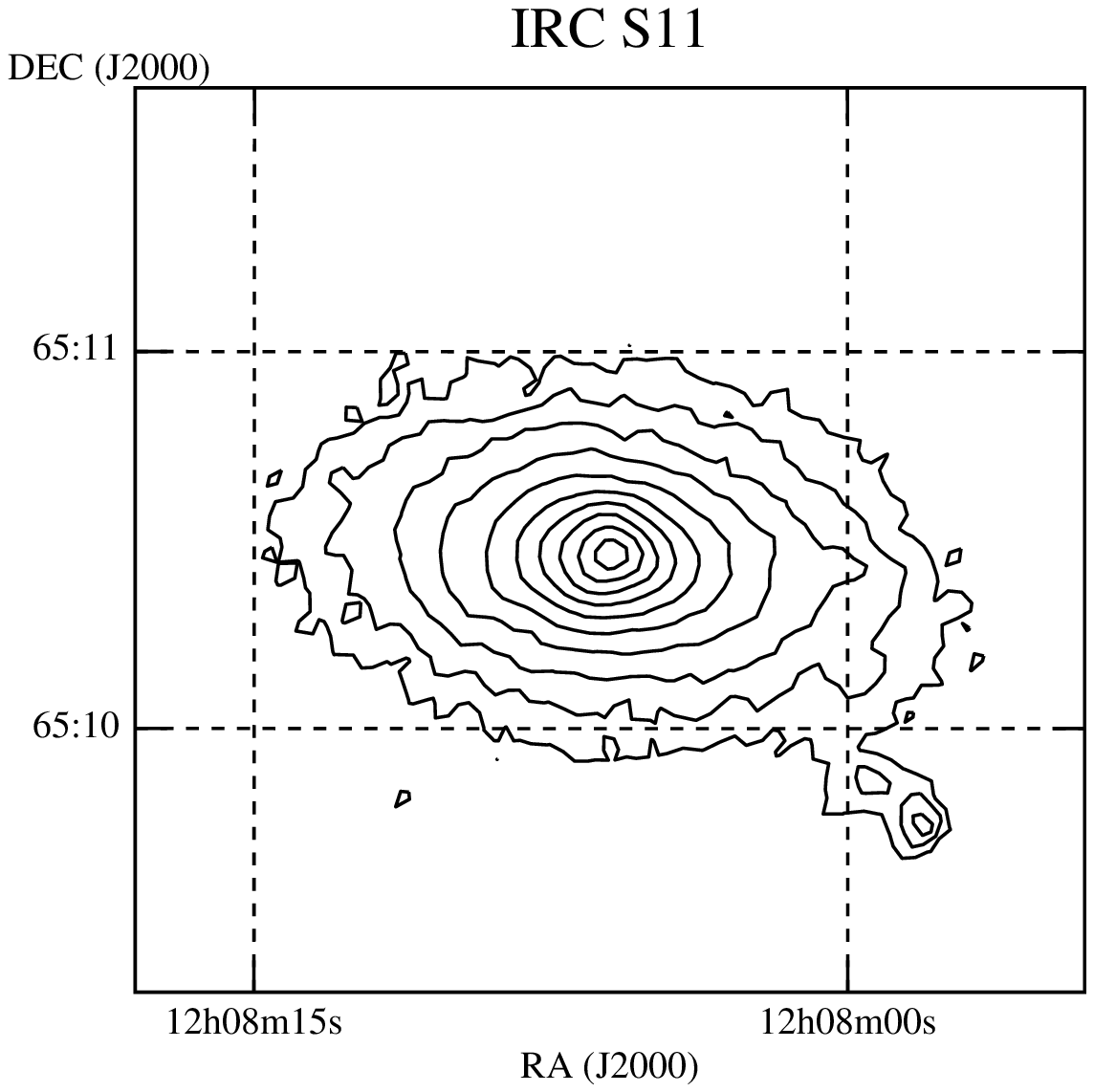}\FigureFile(50mm,50mm){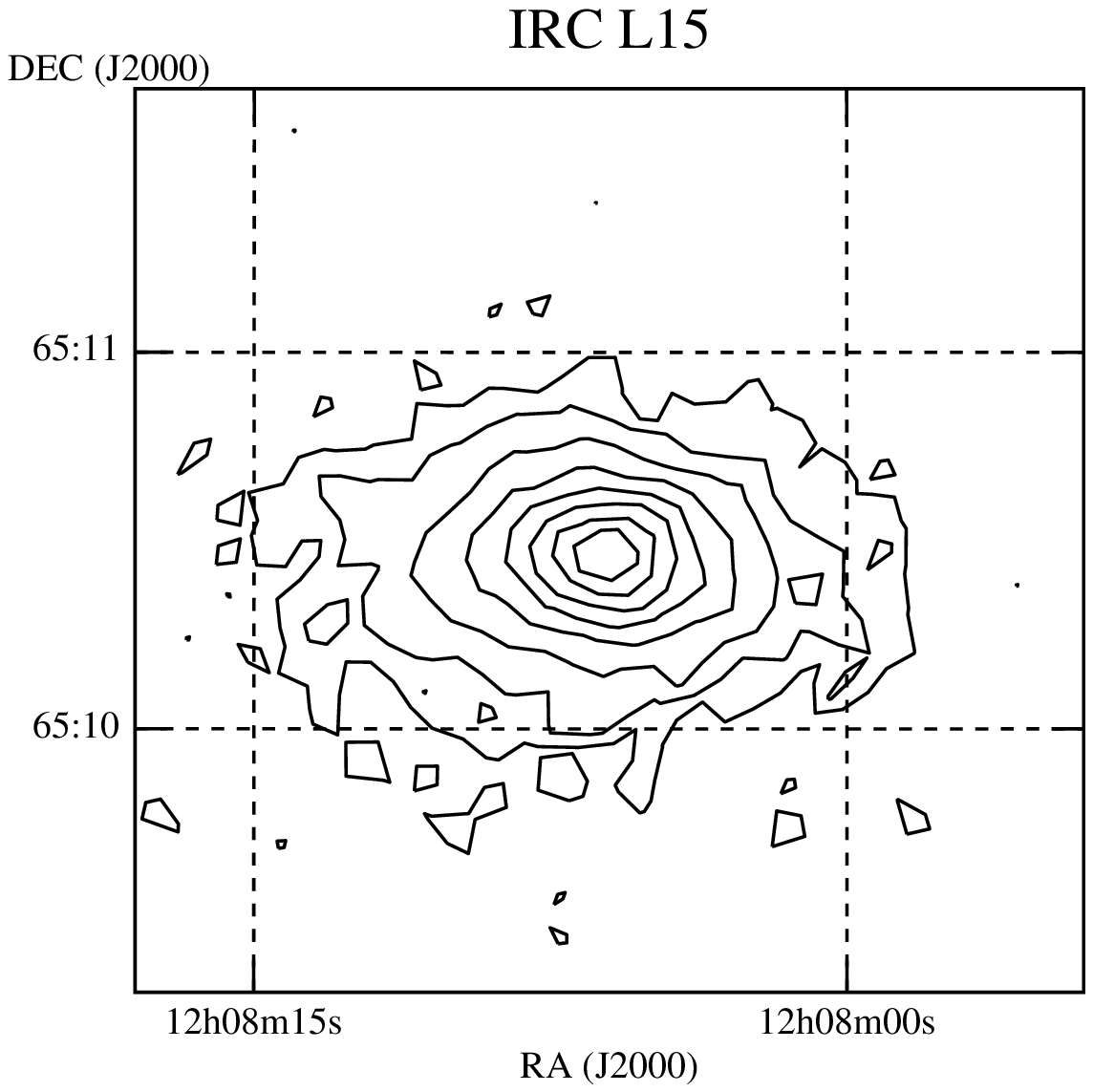}\FigureFile(50mm,50mm){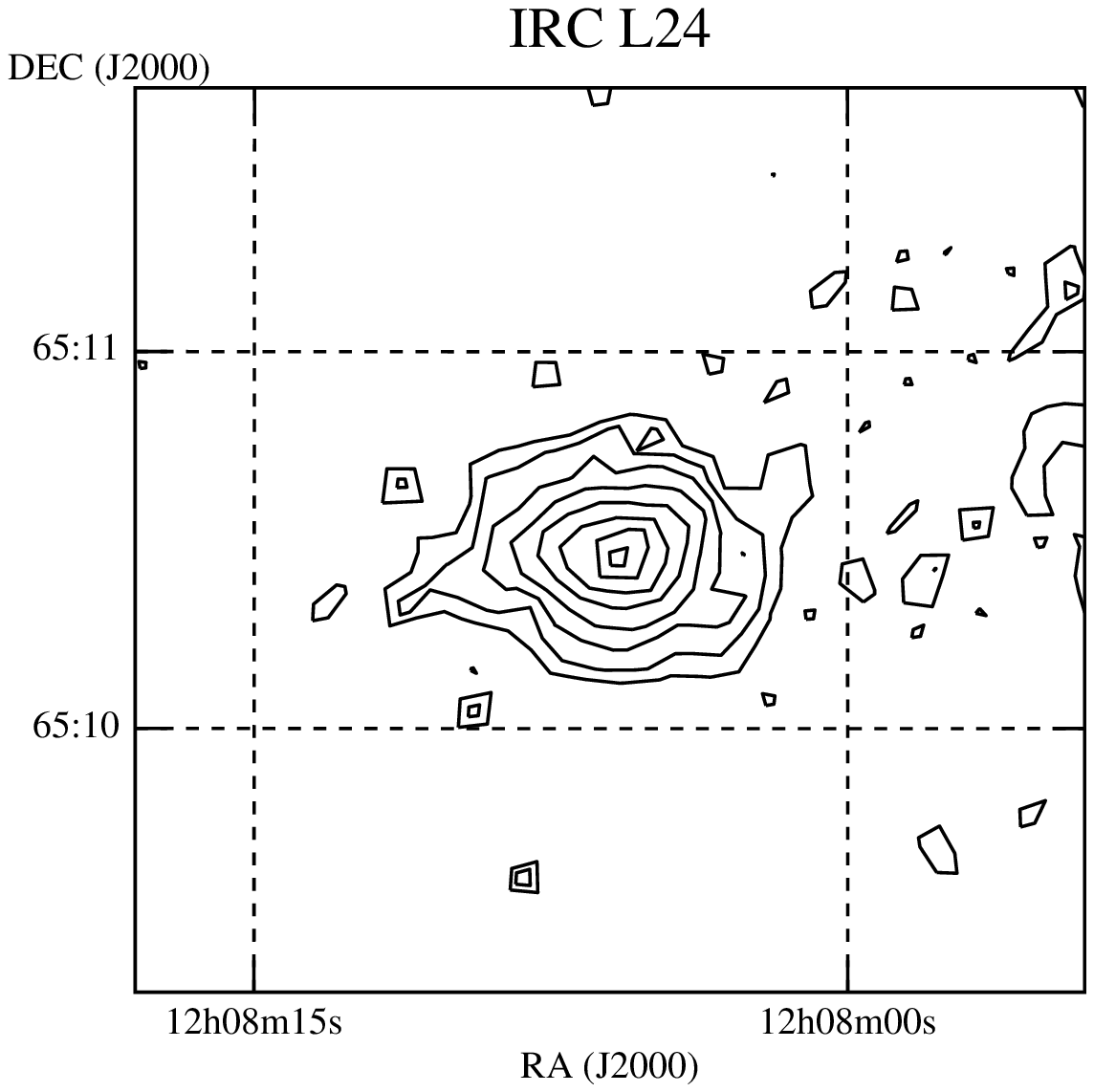}\\
\FigureFile(50mm,50mm){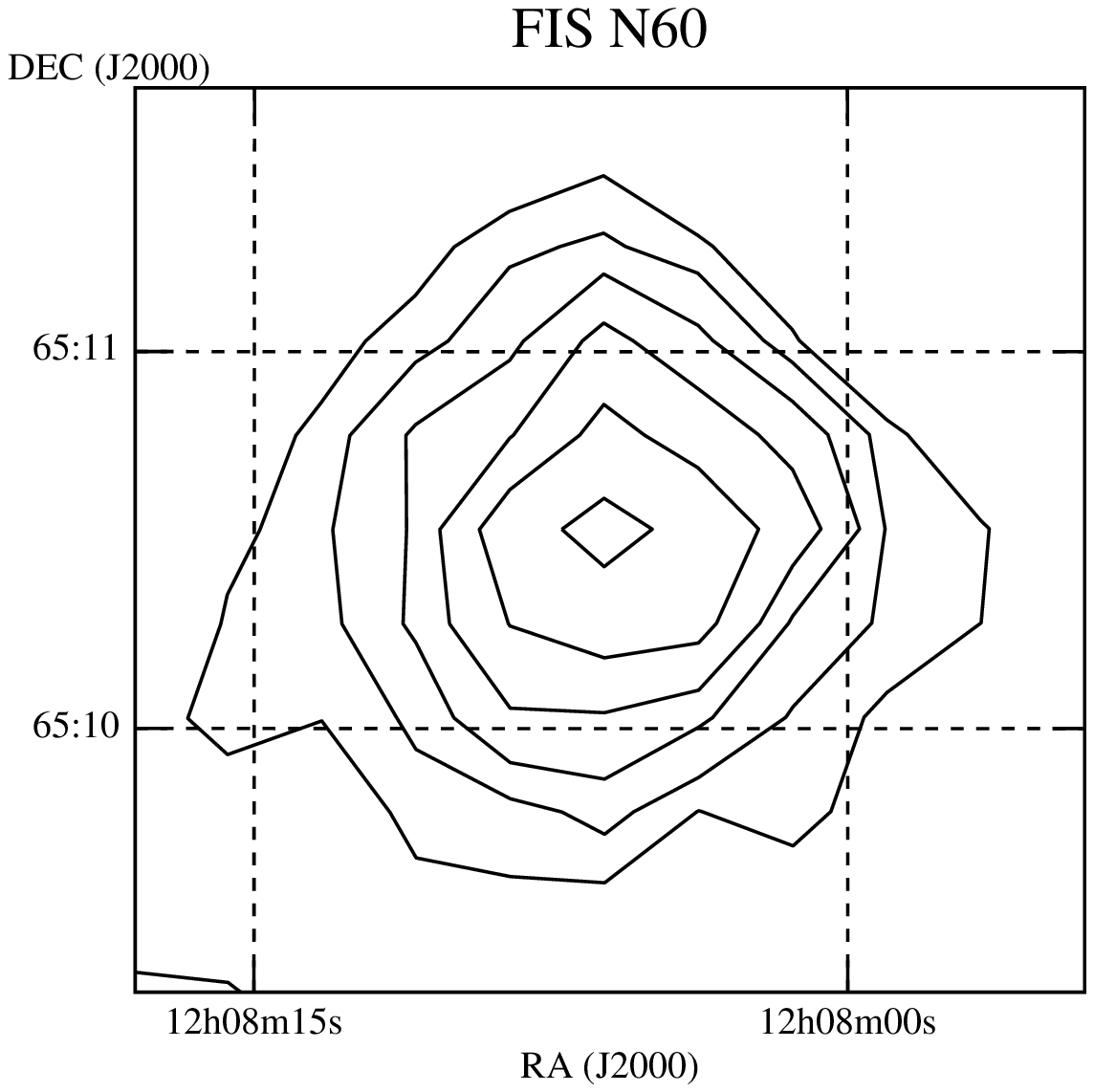}\FigureFile(50mm,50mm){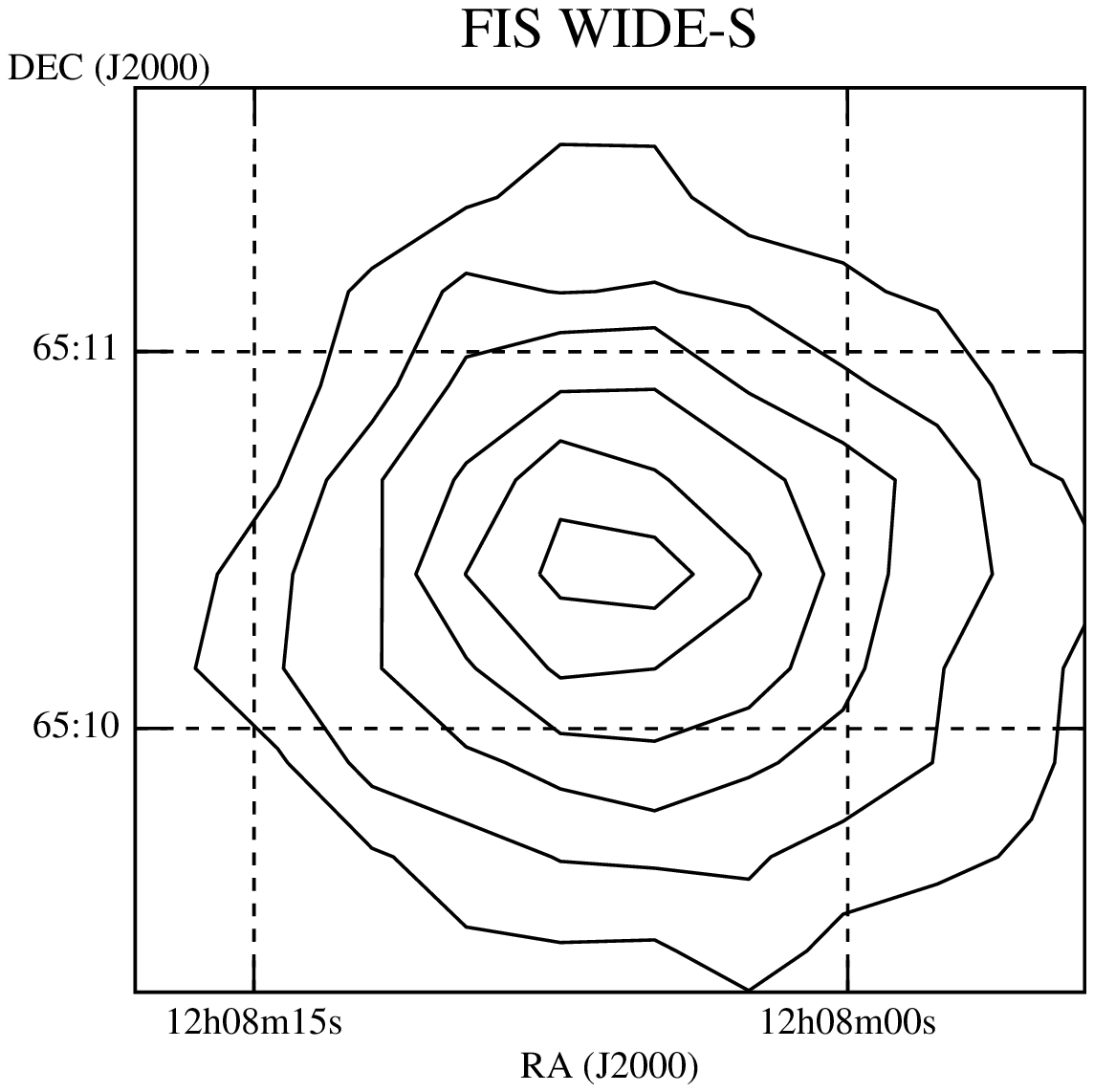}\FigureFile(50mm,50mm){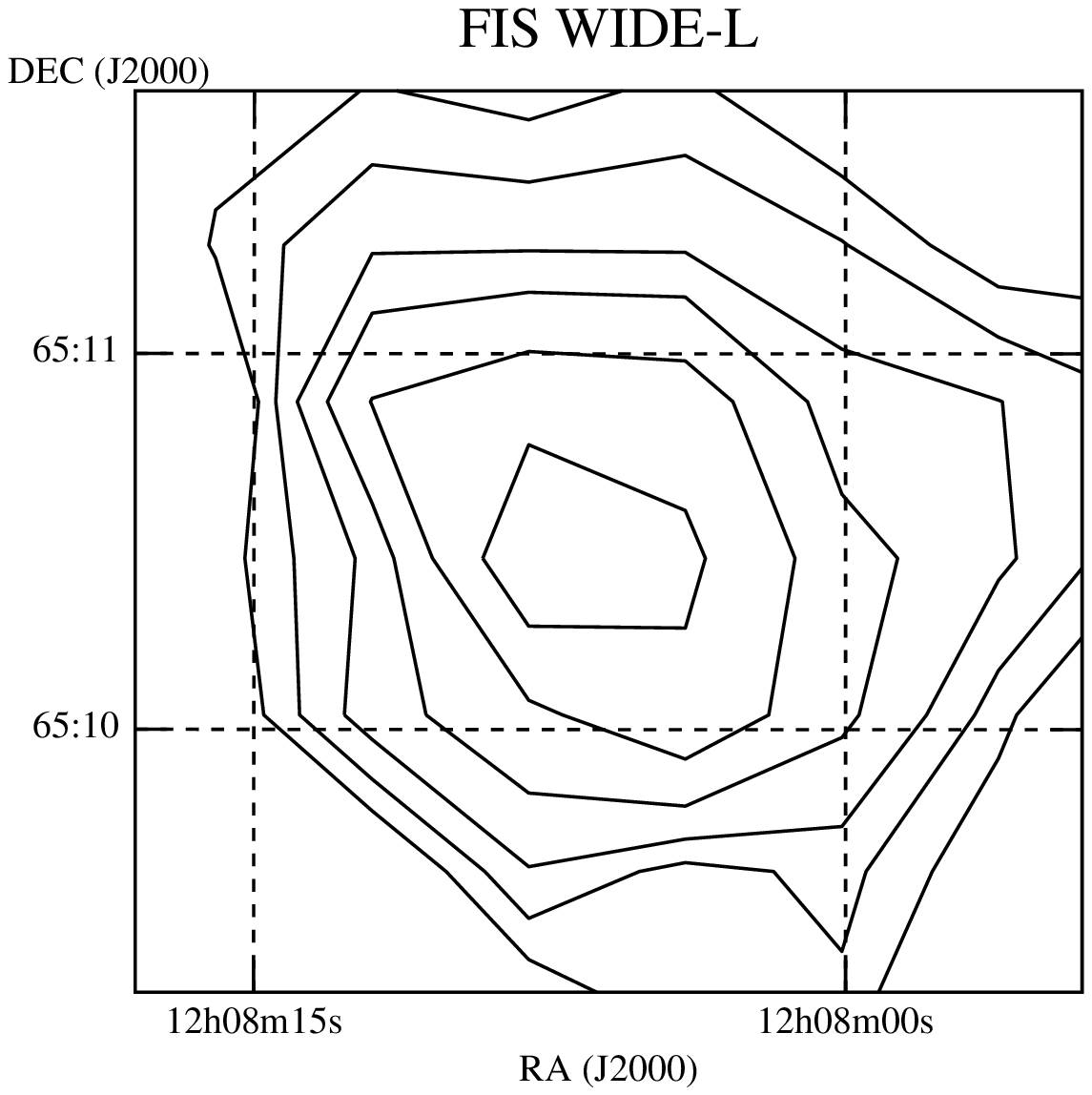}\\
\caption{AKARI/FIS and IRC images of NGC~4125. The contours are drawn at logarithmically-spaced 10 levels from 80 \% to 2 \% of the peak surface brightness. The contour levels below 3 \%, 5 \%, and 10 \% are not shown for the {\it L15}, {\it L24}, and the FIS images, respectively.}
\end{figure}

The {\it N60}, {\it WIDE-S}, and {\it WIDE-L} images, again, show apparent deviations from the smooth stellar distribution as seen in the {\it N3} and {\it N4} bands. 
The {\it WIDE-S} and {\it WIDE-L} images exhibit an apparent elongation to a similar direction to the [NeII] and [SiII] line maps in figure 2 (i.e. P.A. $\simeq$ 70$^{\circ}$), although their spatial scales are quite different. As explained in Kaneda et al. (2008b), the {\it WIDE-S} band image is best among the three and even advantageous over MIPS far-IR images to discuss the spatial distribution of faint far-IR emission in elliptical galaxies. In figure 5, we show the encircled energy profile of the far-IR distribution of NGC~4125 in the {\it WIDE-S} band, comparing to that of the point spread function (PSF) in the same band (Shirahata et al. 2009). From the figure, we conclude that the distribution of the far-IR dust emission is significantly extended even with the large PSF of the far-IR band. For comparison, we also show an encircled energy profile for the PAH 11.3 $\mu$m emission in figure 2. It is obvious that the distribution of the PAH emission is by far more compact than the PSF in the far-IR, and that there is a large difference in size of the distribution between the PAHs and far-IR dust.

\begin{figure}
\FigureFile(150mm,100mm){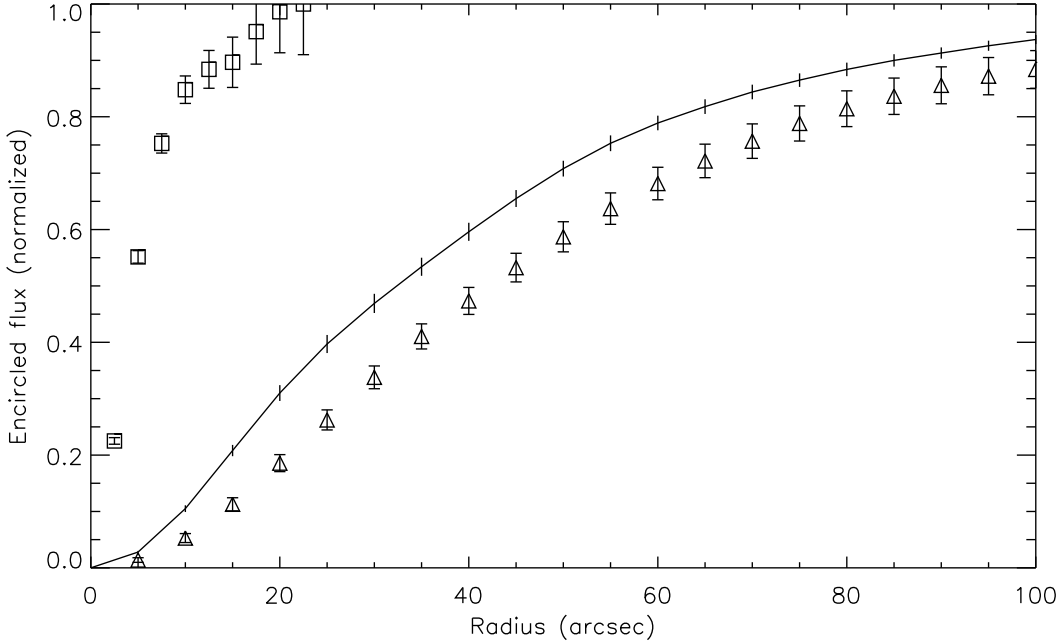}\\
\caption{Encircled energy profile of the far-IR distribution of NGC~4125 in the {\it WIDE-S} band in open triangles, which is compared to that of the standard PSF in the same band in solid curve (Shirahata et al. 2009). The two profiles are normalized to unity at a radius of $180''$. The open squares indicate the encircled energy profile of the distribution of the PAH 11.3 $\mu$m emission. The error bars attached to the open triangles and squares are estimated from sky fluctuations in their image data, while those to the solid curve are the standard deviations of the observed PSFs with different colors (Shirahata et al. 2009).}
\end{figure}

\subsection{Spectral energy distributions and spectral fitting}
We derive the flux densities of NGC~4125 by integrating the surface brightness within the circular aperture radii of $\timeform{14''.1}$ ($\simeq$ 1.5 kpc) and $\timeform{1'.5}$ ($\simeq$ 9.7 kpc) around the galactic center in each AKARI band image. The former aperture has the same area as the largest one for the Spitzer/IRS specta (figure 3). The observed flux densities are listed in table 3, where the errors are estimated from background sky fluctuations.  
For the IRC, no aperture corrections are performed, since the aperture radii are sufficiently large as compared to the PSF sizes. For the FIS, aperture corrections are performed by using the correction tables in Shirahata et al. (2009), while color corrections are not performed because they are found to be negligible considering the systematic errors of the fluxes described below. 

The spectral energy distributions (SEDs) constructed from the flux densities in table 3 are presented in figure 6. For the SED of the central region ($r<\timeform{14''.1}$), the above Spitzer/IRS spectrum of the same area is shown together, while for the SED of the total region ($r<\timeform{1'.5}$), the Spitzer/IRAC and MIPS photometric data points are added (Dale et al. 2007; Temi et al. 2007). For the FIS data points, we consider systematic errors associated with the detectors and absolute uncertainties, which are estimated to be 20 \% for {\it N60} and {\it WIDE-S}, and 30 \% for {\it WIDE-L} (Kawada et al. 2007). The AKARI flux densities in the near- to far-IR show overall agreements with the measurements by Spitzer including both photometric data with the IRAC and MIPS and spectral data with the IRS.

In order to fit the IRS spectra, we have used the PAHFIT tool developed by the Spitzer SINGS legacy team (Smith et al. 2007b). In table 4, the fitting results of the spectra for the $15''\times15''$ and $25''\times25''$ areas are summarized for the intensities of the gas lines, PAH features and continuum emissions that have signal-to-noise ratios higher than 5 for both spectra. To further fit the far-IR photometric data points together with the mid-IR spectra in figure 6, we have changed the temperature of the coolest dust component in the PAHFIT from 35 K to 25 K (figure 6a) or 23 K (figure 6b); all the dust components have the emissivitiy power-law index of $\beta=2$. We have simply extrapolated the best-fit curve in the mid-IR to the near-IR regime. For the total SED in figure 6b, the mid-IR spectral shape is kept the same as the above best-fit curve and only the normalizations are scaled to match the photometry. 

The model reproduces both SEDs well. The {\it S11} data point exceeds the level interpolated by using the other photometric data points, reflecting that the strong PAH 11.3 $\mu$m feature and its underlying plateau emission contribute much to the {\it S11} band intensity. The {\it S7} band intensity is dominated by the stellar continuum emission like the {\it N3} and the {\it N4} band intensity, because usually strong PAH 7.7 $\mu$m band emission is apparently weak; the PAHFIT gives the PAH 7.7 $\mu$m complex intensity of $(5.9\pm1.2)\times10^{-17}$ W m$^{-2}$ for the $15''\times15''$ area and no significant detection ($<1.2\times10^{-16}$ W m$^{-2}$ as a 5-sigma upper limit) for the $25''\times25''$ area. It should be noted, however, that the PAH 7.7 $\mu$m band strength is underestimated unless a stellar silicate feature around 10 $\mu$m is taken into account, and that the difference can be such a large factor as 2--3 for the above feature strength (Kaneda et al. 2008a). Nevertheless, even considering the effect of the stellar component, the PAH7.7/11.3 ratio is 0.6--0.9, which is considerably lower than 3.6, the median, and 1.5--4.8, the 10 \%--90 \% range of variation over the full sample of the normal galaxies in the Spitzer/SINGS legacy program (Smith et al. 2007b). The PAH 17/11.3 ratio is 0.5 (table 4), very similar to the ratio of $\sim$0.5 averaged over all the SINGS sample, suggesting that the size distribution of the PAHs is normal (Draine \& Li 2007). Thus we attribute this unusual weakness of the PAH 7.7 $\mu$m complex intensity to the dominance of neutral PAHs over ionized ones due to very soft interstellar radiation field typical of elliptical galaxies (Kaneda et al. 2005, 2008a). However we detect the PAH 6.2 $\mu$m feature that is apparently broad and strong despite the weak PAH 7.7 $\mu$m feature. Consequently, the PAH6.2/11.3 ratio is 0.8--1.1 (table 3), similar to 1.1 and 0.52--1.5 for the above values of the SINGS sample. By referring to the diagnostic diagram of PAH 6.2 $\mu$m/7.7 $\mu$m versus 11.3 $\mu$m/7.7 $\mu$m ratios in Sales et al. (2010), our result reveals that the 6.2 $\mu$m/7.7 $\mu$m ratio is remarkably large. This might suggest the relative importance of substitution of C by impurity atoms (Hudgins et al. 2005) or clustering of PAHs for the PAH 6.2 $\mu$m feature (Rapacioli et al. 2005; Tielens 2008), although we do not find any other hints from the other PAH features.

\begin{table}
\caption{Observed Flux Densities of NGC~4125}
\begin{center}
\begin{tabular}{lcccc}
\hline\hline
Band name&$\lambda_{\rm ref}$\footnotemark[$*$]&&Center ($r\leq \timeform{14''.1}$)&Total ($r\leq \timeform{1'.5}$)\\
\hline
IRC {\it N3}&3.2 $\mu$m&:&197$\pm$5 mJy&574$\pm$15 mJy\\
IRC {\it N4}&4.1 $\mu$m&:&117$\pm$4 mJy&339$\pm$11 mJy\\
IRC {\it S7}&7.0 $\mu$m&:&61$\pm$2 mJy&178$\pm$4 mJy\\
IRC {\it S11}&11.0 $\mu$m&:&49$\pm$1 mJy&145$\pm$3 mJy\\
IRC {\it L15}&15.0 $\mu$m&:&32$\pm$1 mJy&105$\pm$3 mJy\\
IRC {\it L24}&24.0 $\mu$m&:&26$\pm$2 mJy&80$\pm$5 mJy\\
FIS {\it N60}&65 $\mu$m&:&361$\pm$47 mJy&575$\pm$34 mJy\\
FIS {\it WIDE-S}&90 $\mu$m&:&525$\pm$58 mJy&953$\pm$38 mJy\\
FIS {\it WIDE-L}&140 $\mu$m&:&573$\pm$139 mJy&1250$\pm$83 mJy\\
\hline
\\
\multicolumn{5}{@{}l@{}}{\hbox to 0pt{\parbox{85mm}{\footnotesize

\par\noindent
\footnotemark[$*$] Reference wavelengths (Onaka et al. 2007; Kawada et al. 2007)  
}\hss}}

\end{tabular}
\end{center}
\end{table}

\begin{figure}
\FigureFile(140mm,100mm){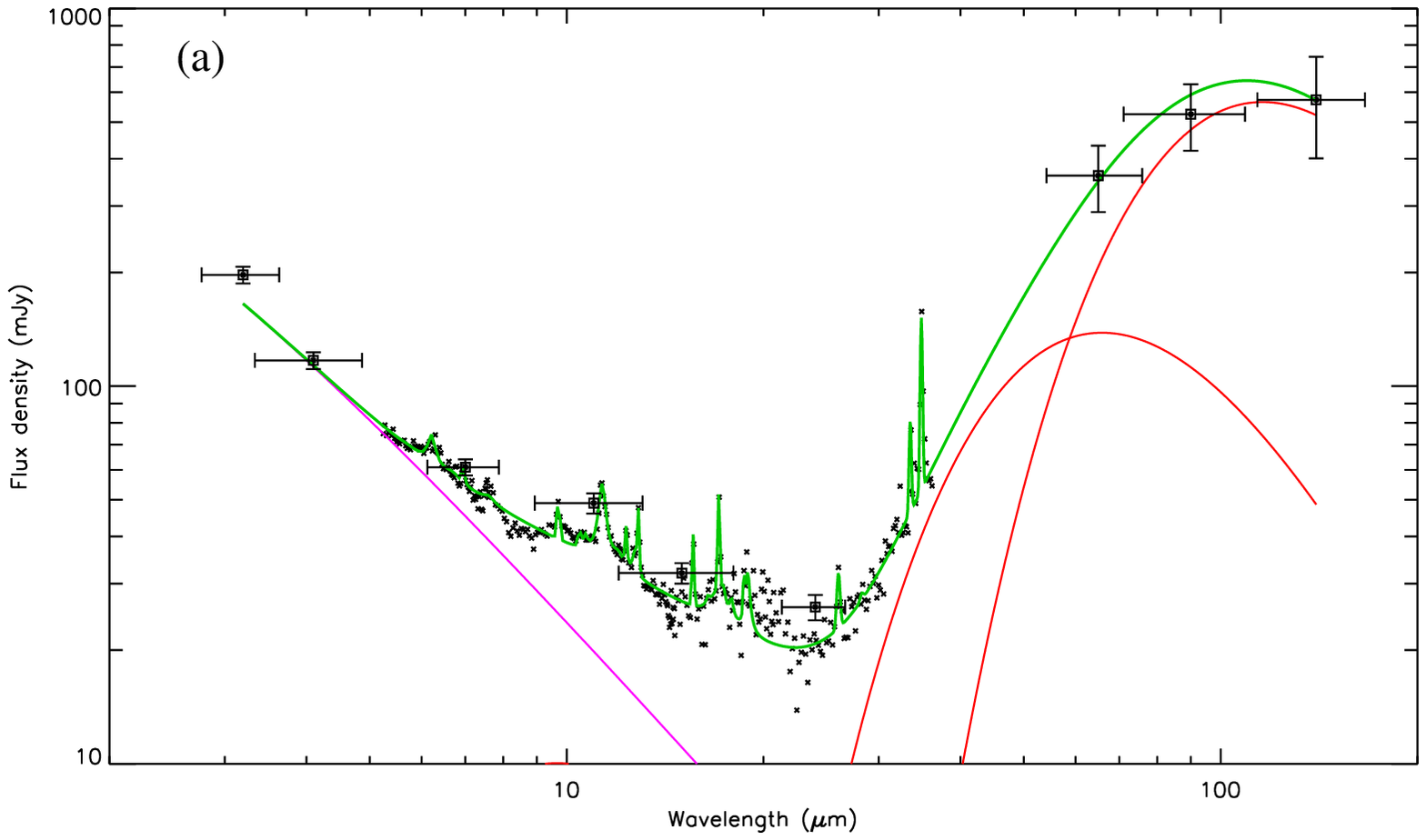}\\
\FigureFile(140mm,100mm){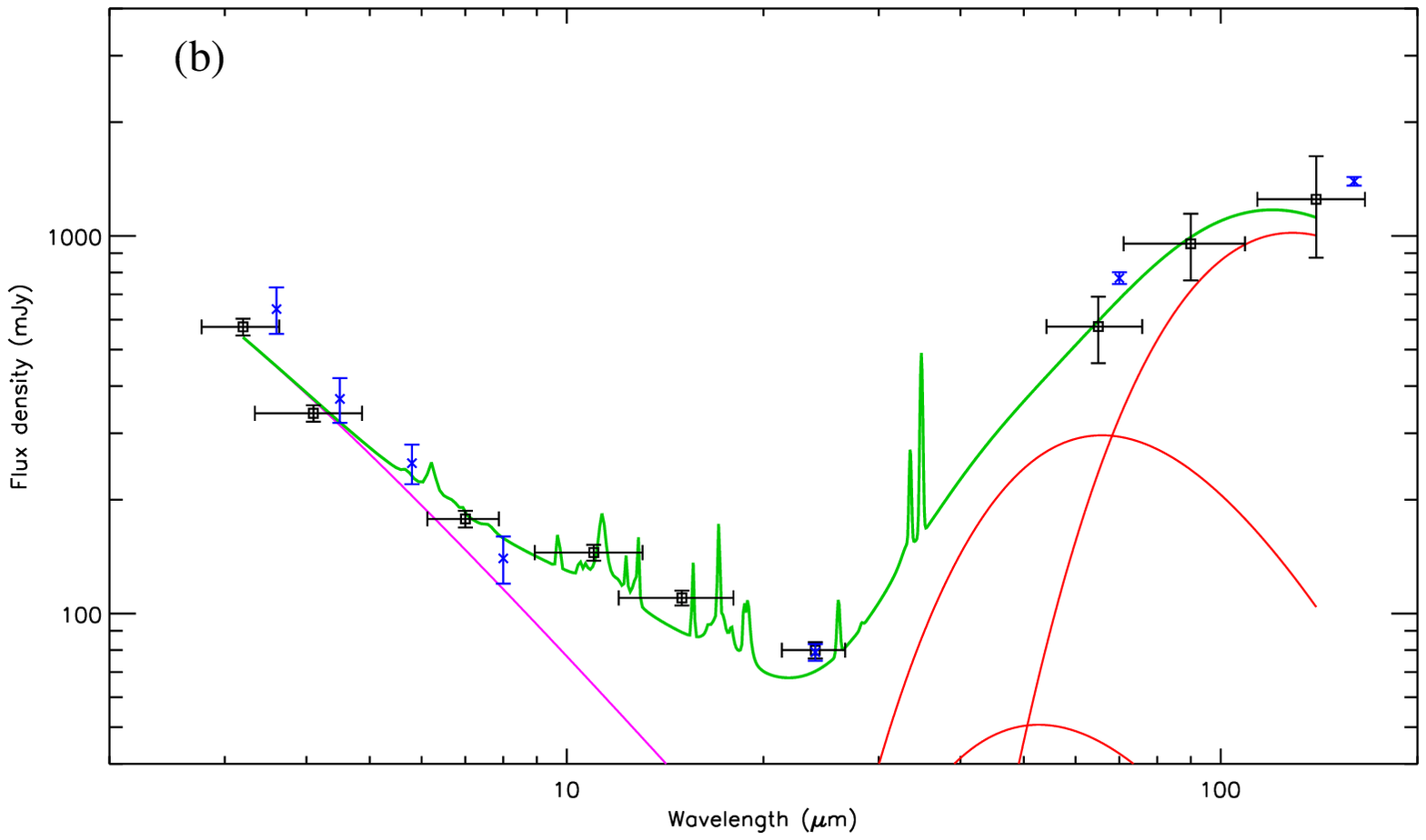}\\
\caption{(a) AKARI photometric data points (boxes with the error bars) for a circular integration aperture of $r=\timeform{14''.1}$ together with the Spitzer spectral data points (with the dots) for the same integration area. The error bars in the wavelength indicate the effective band widths. The green curve represents the curve fitted by the PAHFIT program with a newly-added cooler dust component. The contribution by dust gray-body components with different temperatures is shown by the red curves, while the stellar emission component is shown by the magenta curve. (b) AKARI photometric data points for a circular integration aperture of $r=\timeform{1'.5}$  fitted by the above model where the mid-IR spectral shape is kept the same as in (a) but only the normalization is scaled to match the photometry. The crosses with error bars only in the flux correspond to the Spitzer photometric data points (Dale et al. 2007; Temi et al. 2007).     
}
\end{figure}

\subsection{Physical parameters derived from spectral fitting}
First we obtain a robust estimate on the far-IR (50--170 $\mu$m) luminosity, $L_{\rm AKARI}$, since the FIS {\it N60}, {\it WIDE-S}, and {\it WIDE-L} bands continuously cover the wavelength range from 50 $\mu$m to 170 $\mu$m (Hirashita et al. 2007). The results are $L_{\rm AKARI}=2.8\times10^{8} L_{\odot}$ for $r<\timeform{14''.1}$ and $L_{\rm AKARI}=5.1\times10^{8} L_{\odot}$ for $r<\timeform{1'.5}$. 

We calculate the dust mass, $M_{\rm dust}$, by using the equation (e.g. Hildebrand 1983):
\begin{equation}
M_{\rm dust}=\frac{D^2F_{\nu}}{\kappa_{\nu}B_{\nu}(T)},
\end{equation}
where $D$, $F_{\nu}$, $\kappa_{\nu}$, and $B_{\nu}(T)$ are the galaxy distance, the observed flux density, a dust mass absorption coefficient, and the value of the Planck function at the frequency of $\nu$ and the dust temperature of $T$, respectively. 
We use the 90 $\mu$m flux density and the dust temperature, both given by each dust component in figure 6. We adopt a dust mass absorption coefficient of 36 cm$^2$g$^{-1}$ at 90 $\mu$m given by Hildebrand (1983) and adjusted for $\beta=2$. The dust mass thus derived is $3.4\times10^{5}$ $M_{\odot}$ for $r<\timeform{14''.1}$ and $1.0\times10^{6}$ $M_{\odot}$ for $r<\timeform{1'.5}$. Considering the difference in adopted galaxy distance, the latter dust mass is $\sim3$ times larger than that derived with IRAS by Goudfrooij \& de Jong (1995), while it is $\sim2.5$ times smaller than the mass derived with ISO by Temi et al. (2004) who included a cooler (20 K) dust component. Even colder dust is, however, unlikely to exist predominantly because the IRAM observation of NGC~4125 at 1250 $\mu$m gave 3.5$\pm$1.9 mJy, only a tentative detection (Wiklind \& Henkel 1995). Thus the absolute values of the dust masses are quite uncertain, but the relative change from the inner to the outer region suggests that a significant fraction of dust is distributed over the galaxy.

From the pair of the ortho-H$_2$ S(1) and S(3) line intensities in table 4, we calculate the temperature and column density of the molecular hydrogen gas by using an excitation diagram with the transition probabilities of the H$_2$ pure rotational lines from Black \& Dalgarno (1976). The results are $T=300$ K and $N_{\rm H_2} = 4.1\times 10^{18}$ cm$^{-2}$ for the $15''\times15''$ area and $T=310$ K and $N_{\rm H_2} = 1.7\times 10^{18}$ cm$^{-2}$ for the $25''\times25''$ area. Thus the warm H$_2$ masses are $1.7\times10^{5}$ $M_{\odot}$ for the former and $2.0\times10^{5}$ $M_{\odot}$ for the latter, which are comparable to the dust mass. 
The cold H$_2$ mass is estimated to be $7\times10^{7}$ $M_{\odot}$ from tentative CO detection by Wiklind et al. (1995), which gives the gas-to-dust mass ratio of 70--200, similar to the accepted value of 100--200 for our Galaxy (Sodroski et al. 1997).

On the other hand, the line pair of the [SIII] 33 $\mu$m and 19 $\mu$m lines is a good measure of an electron density $n_e$ of ionized gas. From table 4, their intensity ratios correspond to $n_e\simeq350$ for the $15''\times15''$ area and $n_e\lesssim30$ for the $25''\times25''$ area (Rubin et al. 1994). 
The highly-ionized gas responsible for the lines such as [SIII], [NeII], and [NeIII] is likely to be associated with the LINER nucleus \citep{Smi07b}. Alternatively, the highly-ionized gas emission might be powered by interaction with the hot plasma through heat conduction (Sparks et al. 1989). Because the [NeII] line emission is extended in similar directions to the X-ray plasma (see below), we prefer the latter scenario. As clearly demonstrated in figure 3b, the [NeII] line emission is significantly more extended than the PAH 11.3 $\mu$m emission.  

As for the relative intensities of the dust, molecular, and atomic gas emissions, the ratio of the [SiII] line intensity to the total AKARI far-IR flux is $4.4\times10^{-3}$ for the aperture of the same area ($25''\times25''$ or $r<\timeform{14.1''}$), which is considerably higher than $2\times10^{-3}$, the average of the SINGS sample galaxies (Roussel et al. 2007). The ratio of the H$_2$ lines (S(1)--S(3)) to the [SiII] line intensity for the central $15''\times15''$ area is also as high as $\sim$1.5, which is well outside the range of 0.15--0.5 for the nuclei of the SINGS sample galaxies (Roussel et al. 2007). Hence the H$_2$ line emissions are extremely bright and the [SiII] line is also strong relative to the far-IR emission, as compared to the canonical relation for the normal galaxies, suggesting that they do not originate simply from photo-dissociated regions. There might be a dominant contribution from slow shock heating by dynamical perturbation of gas clouds in the center for the former (Roussel et al. 2007) and Si enrichment through sputtering destruction of dust by hot plasma for the latter (Kaneda et al. 2008a). The strong [SiII] line may also be due to the cooling of X-ray dominated regions around the LINER nucleus (Dale et al. 2006).

\begin{table}
\caption{Intensities of gas lines, PAH features, and continuum emissions}
\begin{center}
\begin{tabular}{lrrrrrrr}
\hline\hline
Area&[NeII]&[NeIII]&[SIII]&[SIII]&[SiII]&H$_2$S(1)&H$_2$S(3)\\
    &12.82 $\mu$m&15.55 $\mu$m&18.71 $\mu$m&33.45 $\mu$m&34.77 $\mu$m&17.04 $\mu$m&9.66 $\mu$m\\
\hline
$15''\times15''$&2.36$\pm$0.10\footnotemark[$*$]&2.04$\pm$0.08&0.88$\pm$0.09&1.11$\pm$0.12&4.55$\pm$0.18&3.24$\pm$0.15&2.28$\pm$0.09\\
$25''\times25''$&3.08$\pm$0.17&2.52$\pm$0.21&1.23$\pm$0.15&2.35$\pm$0.24&7.93$\pm$0.26&4.00$\pm$0.26&3.35$\pm$0.15\\
\hline\hline
Area&\multicolumn{4}{c}{PAH}&\multicolumn{3}{c}{Continuum}\\  
    &6.2 $\mu$m&11.3 $\mu$m&12.7 $\mu$m&17 $\mu$m&6 $\mu$m&15 $\mu$m&34 $\mu$m\\ 
\hline
$15''\times15''$&16.3$\pm$1.2&19.81$\pm$0.39&4.61$\pm$0.85&10.39$\pm$0.66&42.74$\pm$0.33&15.49$\pm$0.45&31.4$\pm$1.4\\
$25''\times25''$&26.5$\pm$2.3&24.25$\pm$0.63&5.77$\pm$0.50&12.7$\pm$1.3&68.27$\pm$0.53&27.3$\pm$1.1&57.3$\pm$3.3\\
\hline
\\
\multicolumn{8}{@{}l@{}}{\hbox to 0pt{\parbox{170mm}{\footnotesize

\par\noindent
\footnotemark[$*$]  The values are given in units of mJy for the continuum emissions and $10^{-17}$ W m$^{-2}$ for the feature and the line emissions. 
}\hss}}

\end{tabular}
\end{center}
\end{table}



\section{Discussion}
\subsection{PAHs in the central dense gas region}
We discuss the distribution of the PAH emission. In figure 7a, the $5.5-6.5$ $\mu$m continuum emission shows a smooth stellar distribution, whereas the PAH 11.3 $\mu$m emission exhibits an apparently compact and different distribution from the stars. Since PAHs can be excited even by visible and near-IR soft photons from old stars (Li \& Draine 2002), the compact distribution of the PAH 11.3 $\mu$m emission strongly suggests the real absence of neutral PAHs in the outer regions of the galaxy. It is possible that PAHs are increasingly becoming charged and the 11.3 $\mu$m emission becomes weaker from the center to the outward (Draine \& Li 2001; Li \& Draine 2001). However the PAH 7.7 $\mu$m emission, which traces the distribution of charged PAHs, does not tend to become stronger in the outer regions, as seen in Fig.3a. A more likely interpretation is that PAHs have been destroyed in the X-ray hot plasma outside the central dense gas region, because PAHs are expected to be destroyed in so short a timescale as $10^2$ yr for the plasma parameters typical of X-ray elliptical galaxies (Micelotta et al. 2010; see below). Deep in dense molecular gas, PAHs are shielded and may be efficiently hydrogenated, which can explain the observed strong 11.3 $\mu$m emission.

Figures 7b and 7c show comparisons of the PAH emission with the $V-I$ color index image and the continuum-subtracted H$\alpha$ image, respectively, which are taken from the NASA/IPAC Extragalactic Database. The former image reveals the presence of dense gas regions near the galactic center, and the PAH emission follows its distribution. Verdoes Kleijn \& de Zeeuw (2005) observed a central dust lane of size $\timeform{1''.3}$ with the difference in P.A. of $+13^{\circ}$ from the galaxy major axis. Although the spatial scale is much larger than this, the PAH emission region has a spatial correspondence with the optical dust lane similarly to the case of NGC~4589, where the PAH emission comes predominantly from the optical dust lane of the elliptical galaxy (Kaneda et al. 2010). It is notable that the distribution of the H$\alpha$ emission also resembles those of the dust lane and PAHs. It is likely that the H$\alpha$ emission is not simply related to star formation but rather cooling gas from the hot plasma or heating gas from the dense cold gas through thermal conductance at their contacting interfaces.  

As seen in the spectra in figure 6, the {\it S11} band is expected to contain strong PAH 11.3 $\mu$m emission and its underlying plateau feature probably due to PAH clusters, in addition to the stellar continuum emission. The contribution of the [NeII] line is negligible from table 4. Since the {\it S7} band is dominated by the stellar continuum emission, we can estimate the spatial distributions of the PAHs and PAH clusters emitting the features around 11 $\mu$m by normalizing the {\it S7} image properly and subtracting it from the {\it S11} image. We determined the normalization factor on the basis of the stellar continuum predicted by the spectral fitting in figure 6a. 
In the {\it S11}-{\it S7} differential image thus derived (figure 7d), the emission extends along the east-west direction (i.e. P.A.$=90^{\circ}$) rather differently from the major axis P.A. $82.5^{\circ}$ of the stellar distribution, and connects to the inner distribution of the PAH 11.3 $\mu$m emission in figure 7a. The broader distribution than the PAH 11.3 $\mu$m emission is probably attributed to the PAH clusters responsible for the 11 $\mu$m plateau emission. 

\begin{figure}
\FigureFile(80mm,80mm){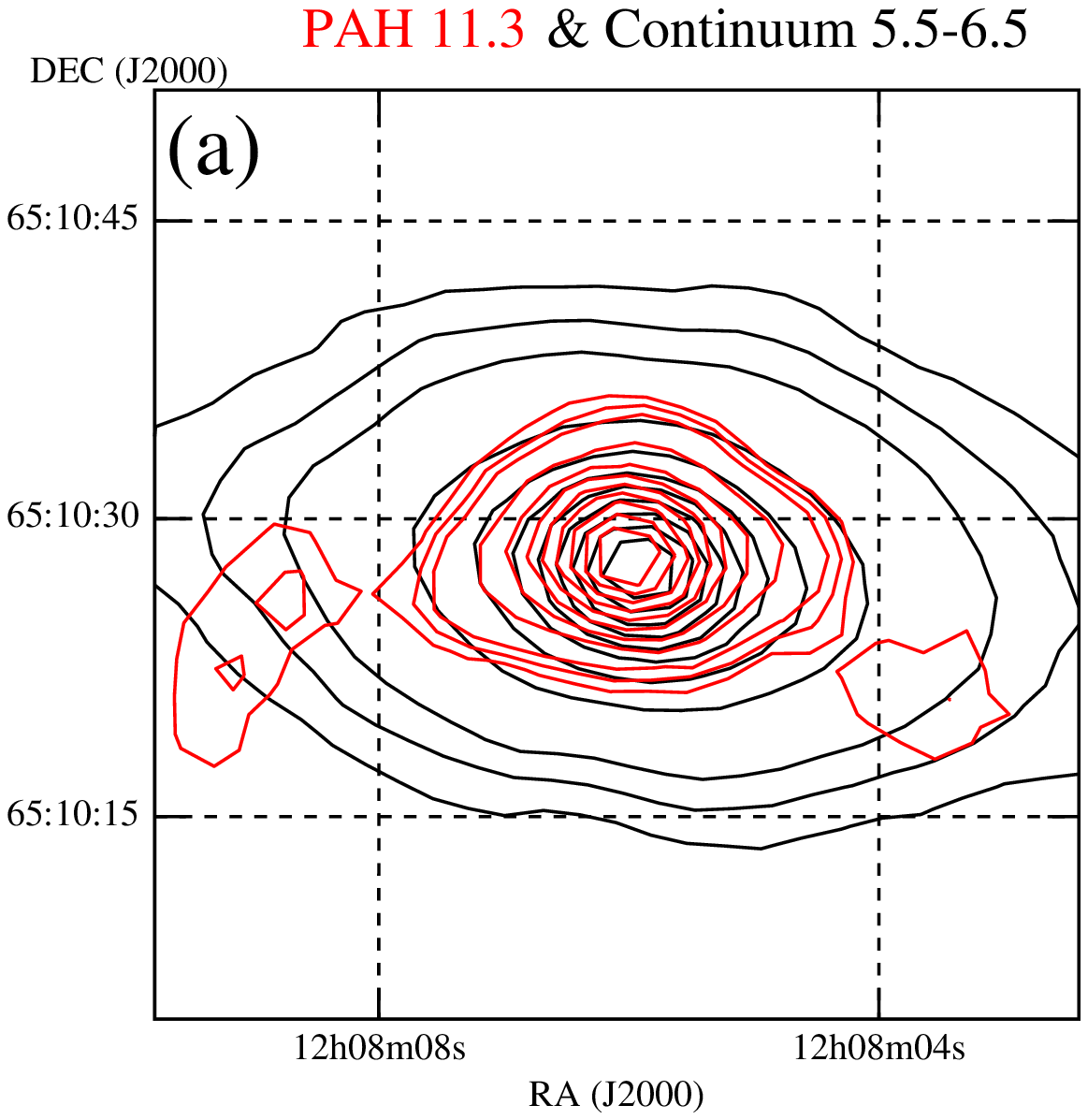}\FigureFile(80mm,80mm){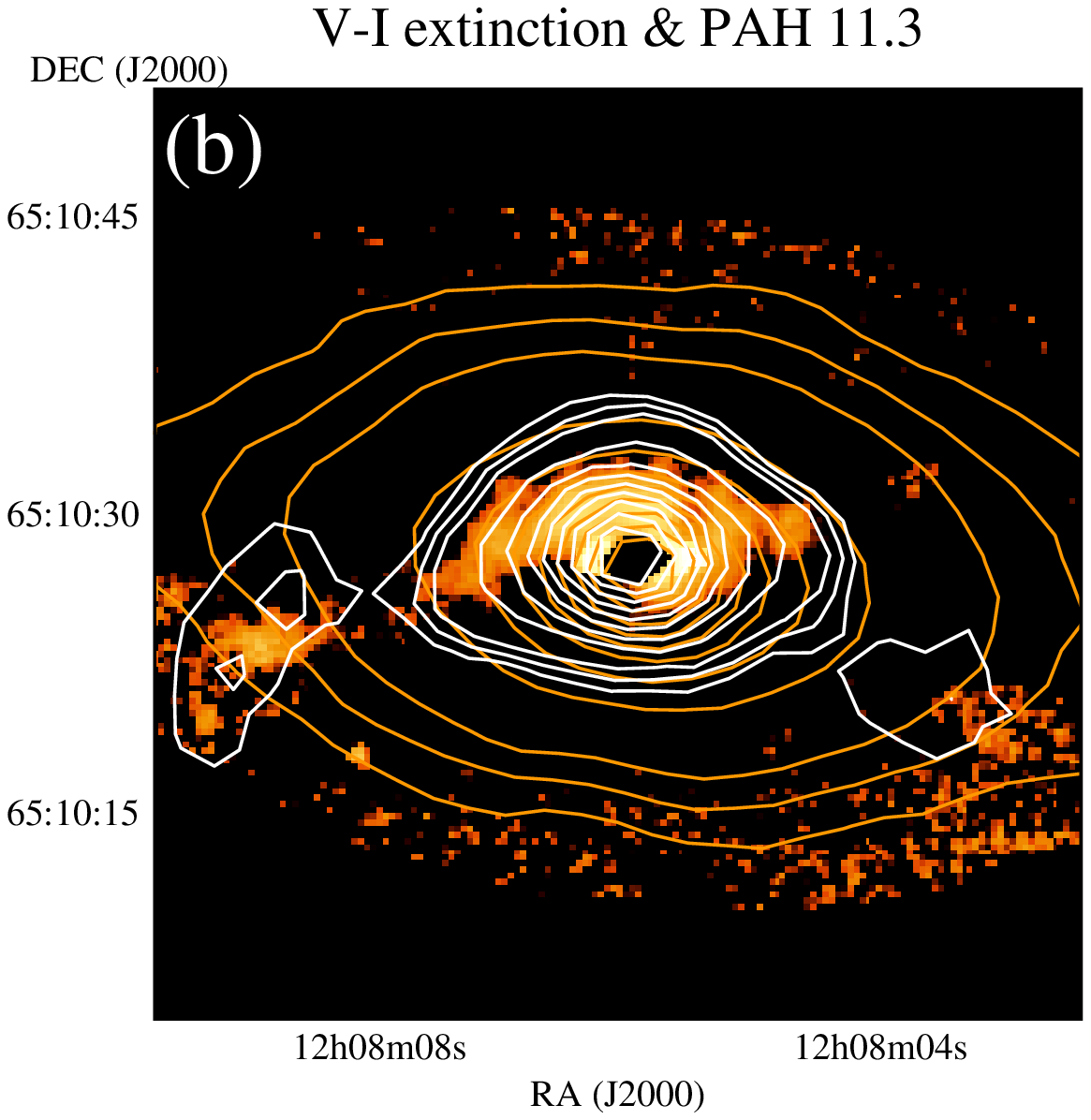}\\
\FigureFile(80mm,80mm){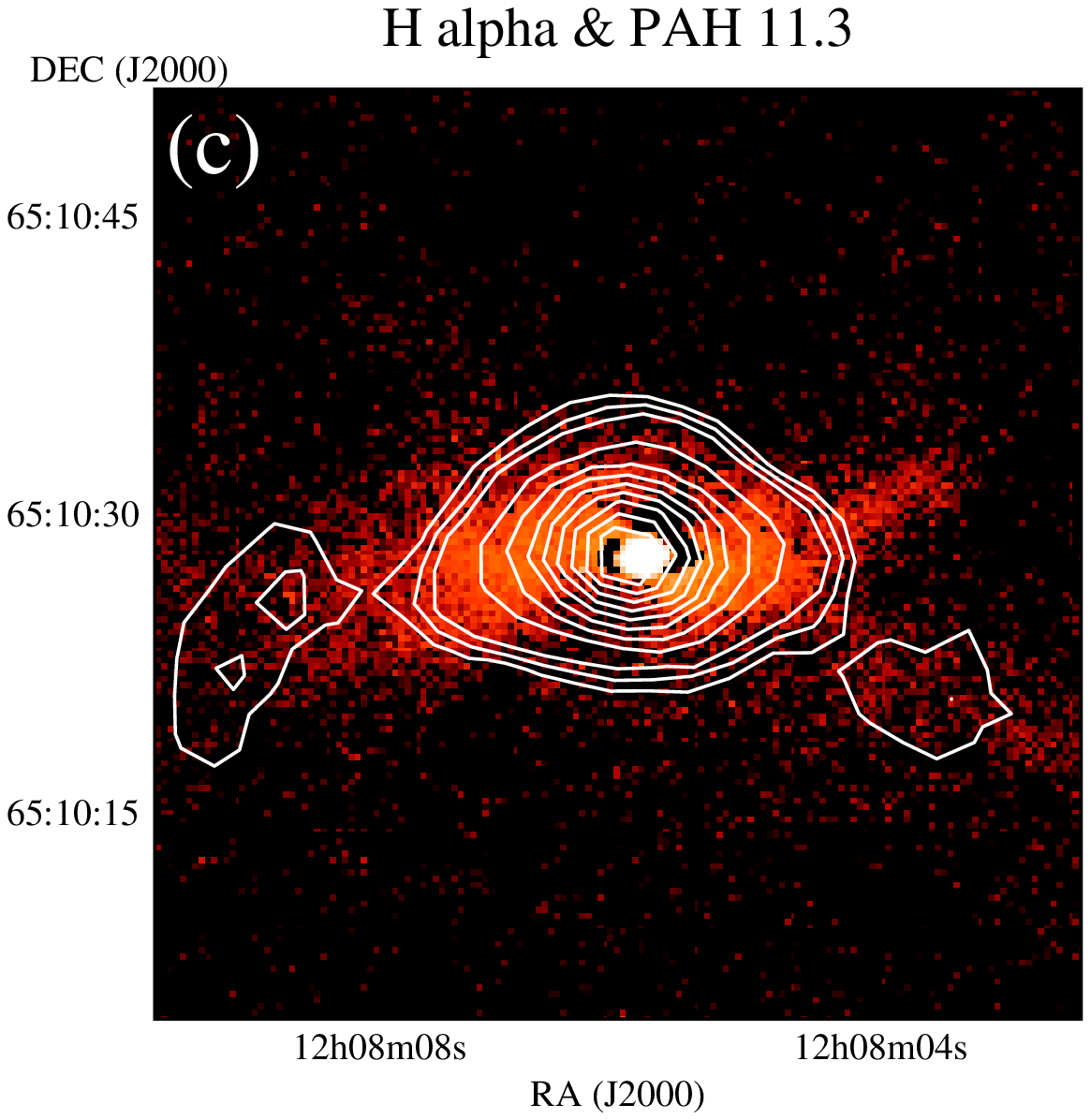}\FigureFile(80mm,80mm){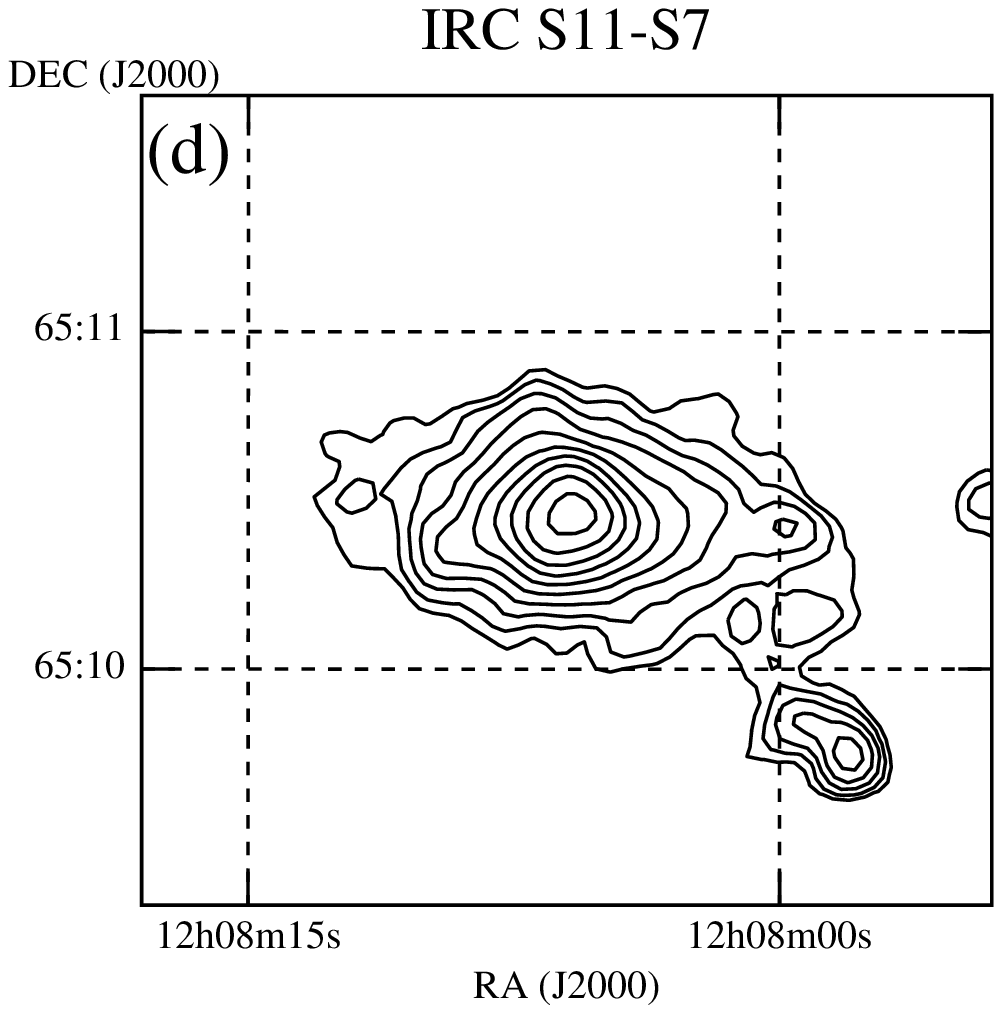}\\
\caption{(a) Spitzer/IRS contour map of the PAH 11.3 $\mu$m feature (red lines) overlaid on that of the $5.5-6.5$ $\mu$m continuum emission (black lines). Both are the same as in figure 2. (b) The $V-I$ color index image superposed on the same contour maps as in the panel (a). (c) The continuum-subtracted H$\alpha$ image superposed on the same PAH 11.3 $\mu$m map as in the panel (a). (d) AKARI/IRC {\it S11} image shown after subtraction of the {\it S7} image with its intensity adjusted on the basis of the SED in figure 6a. The contours are drawn at logarithmically-spaced 10 levels from 80 \% to 2 \% of the peak surface brightness. Note that spatial scales are different between the panels (a)--(c) and the panel (d)}
\end{figure}

\subsection{Dust with diffuse hot plasma and atomic gas}
We compare the distribution of dust with those of the X-ray plasma and the atomic gas.
Figure 8a shows the X-ray (0.3--7 keV) image retrieved from the Chandra data archive.
In order to obtain the spatial distribution of diffuse hot plasma alone, X-ray point sources that are thought to be low-mass X-ray binaries were removed from the image by using the Chandra Interactive Analysis of Observations (CIAO) software. The X-ray image was then smoothed and the contours were superposed on the original image. As seen in figure 8a, NGC~4125 certainly possesses diffuse hot plasma, but its distribution is rather different from the stellar distribution. Figure 8b shows a comparison between the X-ray plasma and the stellar continuum emission in the central $45''\times45''$ area, revealing that the X-ray distribution has a smaller P.A ($\sim$70$^{\circ}$) than the stellar major axis P.A. ($\sim\timeform{82.5^{\circ}}$) of the galaxy. Figures 8c and 8d show the X-ray images in a soft (0.3---0.8 keV) and a hard (0.8--7 keV) band, respectively, which clearly exhibit that cooler plasma is distributed more widely to the direction of P.A.$\sim$70$^{\circ}$. Trinchieri et al. (2000) already pointed out that the ASCA and BeppoSAX 0.2--10 keV spectrum of NGC~4125 cannot be reproduced by a single thermal model, calling for the presence of a second soft (kT$\sim$0.3--0.7 keV) spectral component.

Figures 9a and 9b show the spatial distribution of the far-IR dust compared to those of the atomic gas ([SiII]) and X-ray hot plasma, respectively; they bear resemblances to each other, which are extended to similar directions of P.A.$\sim$70$^{\circ}$. The spatial correlation in figure 9a may suggest relatively abundant gas-phase Si through plasma destruction of dust where Si is depleted. As shown in figure 9b, the X-ray hot plamsa is obviously of a diffuse nature, and the dust emission is distributed similarly; such a relationship is rather unexpected because the lifetime of dust with sizes of 0.1 $\mu$m exposed to hot plasma with gas densities of $10^{-2}\sim 10^{-3}$ cm$^{-3}$ is quite short ($10^6\sim 10^7$ yr; Draine \& Salpeter 1979; Tielens et al. 1994; Tsai \& Mathews 1995). NGC~4125 has electron densities and temperatures of $(2-10)\times10^{-3}$ cm$^{-3}$ and 0.36--0.44 keV at 3--10 kpc radii (Fukazawa et al. 2006). The supply of the dust by old stars as dominant sources is unlikely considering the clear difference in distribution between the dust and stars. 
If the observed dust mostly originates from external sources, their lifetime is obviously shorter than  the crossing time of a galaxy through a merging galaxy ($\sim 10^8$ yr), where we assume the galaxy size of 10 kpc diameter with an infall velocity of 100 km s$^{-1}$. 
A possible explanation is the evaporation flow scenario (Sparks et al. 1989; de Jong et al. 1990), 
where evaporating clouds were brought in by a gas-rich galaxy that merged in the past. The evaporation of the clouds is due to thermal conduction by plasma eletrons resulting in cooling the X-ray plasma. For NGC~4696, which has X-ray plasma parameters similar to the above, de Jong et al. (1990) calculated that clouds with column densities of $N_{\rm H}\simeq 7\times10^{21}$ cm$^{-2}$ or $A_V\simeq 3$ can survive in $10^8$ yr. The alignment of the distributions of the dust to the softer plasma emission supports the evaporation cloud scenario. In the present case, however, clouds must be sufficiently small, and thus dense, because they are not observed in the extinction map (figure 7b). Even if clouds escape detection in the optical absorption, they would have to be observed in the PAH emission. However, the observed PAH emission does not have such extended components, either. In fact, the ratio of the integrated PAH ($6.2-17$ $\mu$m) intensity to the total far-IR flux is $3.5\times10^{-2}$ for the area of $25''\times25''$ (or $r<\timeform{14.1''}$), which is considerably lower than $0.10\pm0.04$ for the full sample of SINGS galaxies (Smith et al. 2007b). Therefore it is difficult to explain the observed dust emission by the presence of such dense clumpy clouds in the outer region of the galaxy.  

\begin{figure}
\FigureFile(80mm,80mm){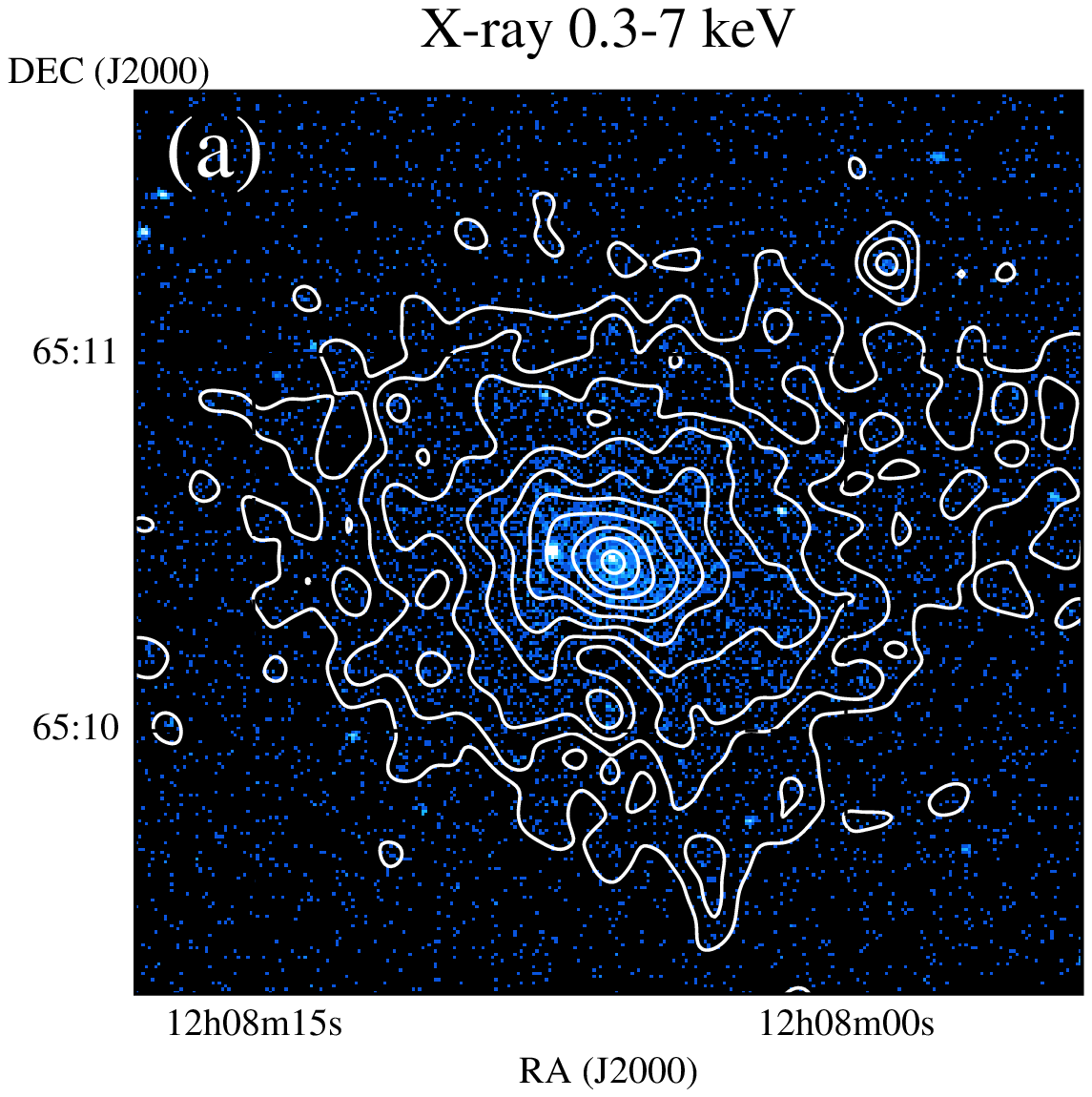}\FigureFile(80mm,80mm){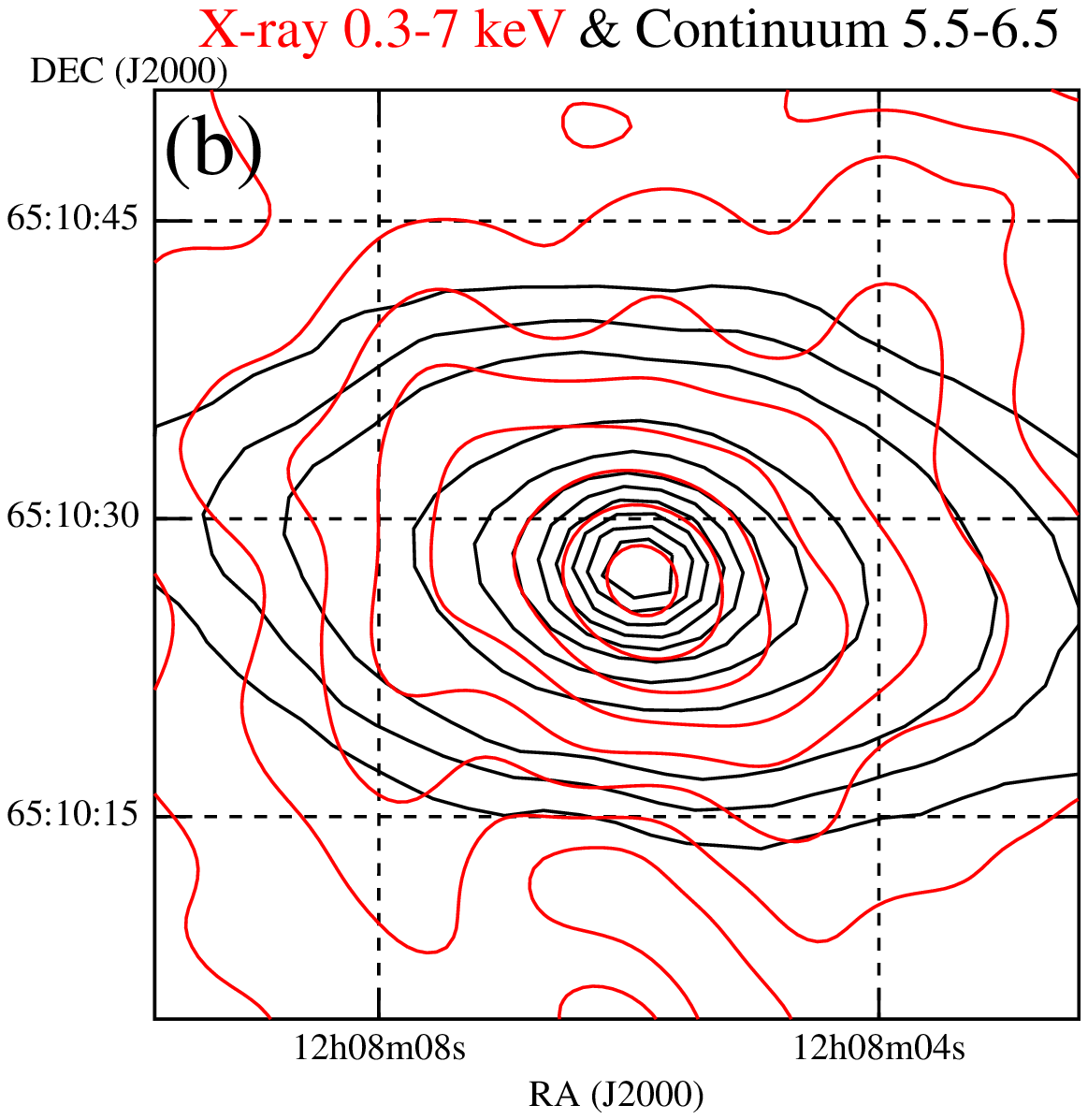}\\
\FigureFile(80mm,80mm){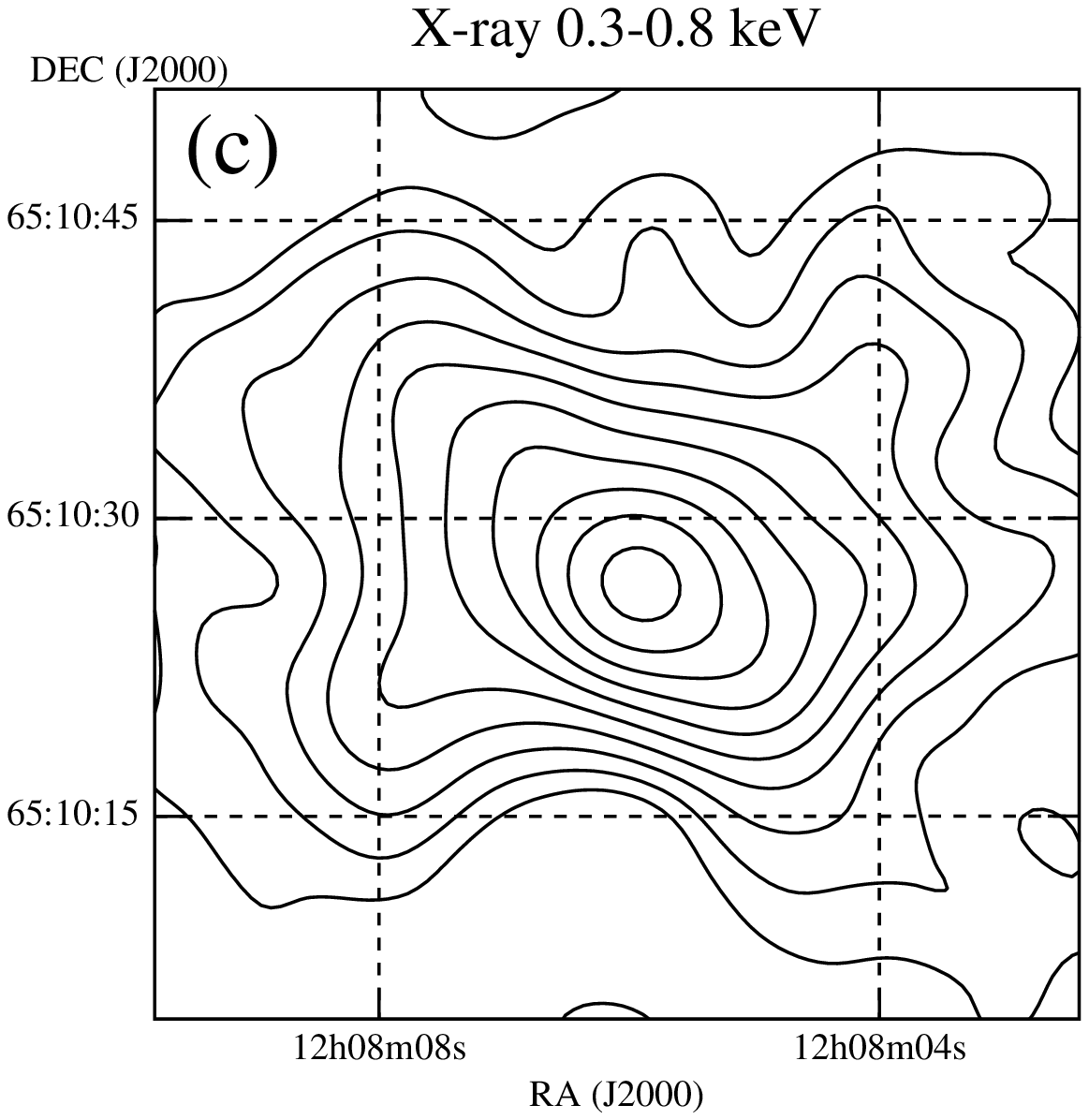}\FigureFile(80mm,80mm){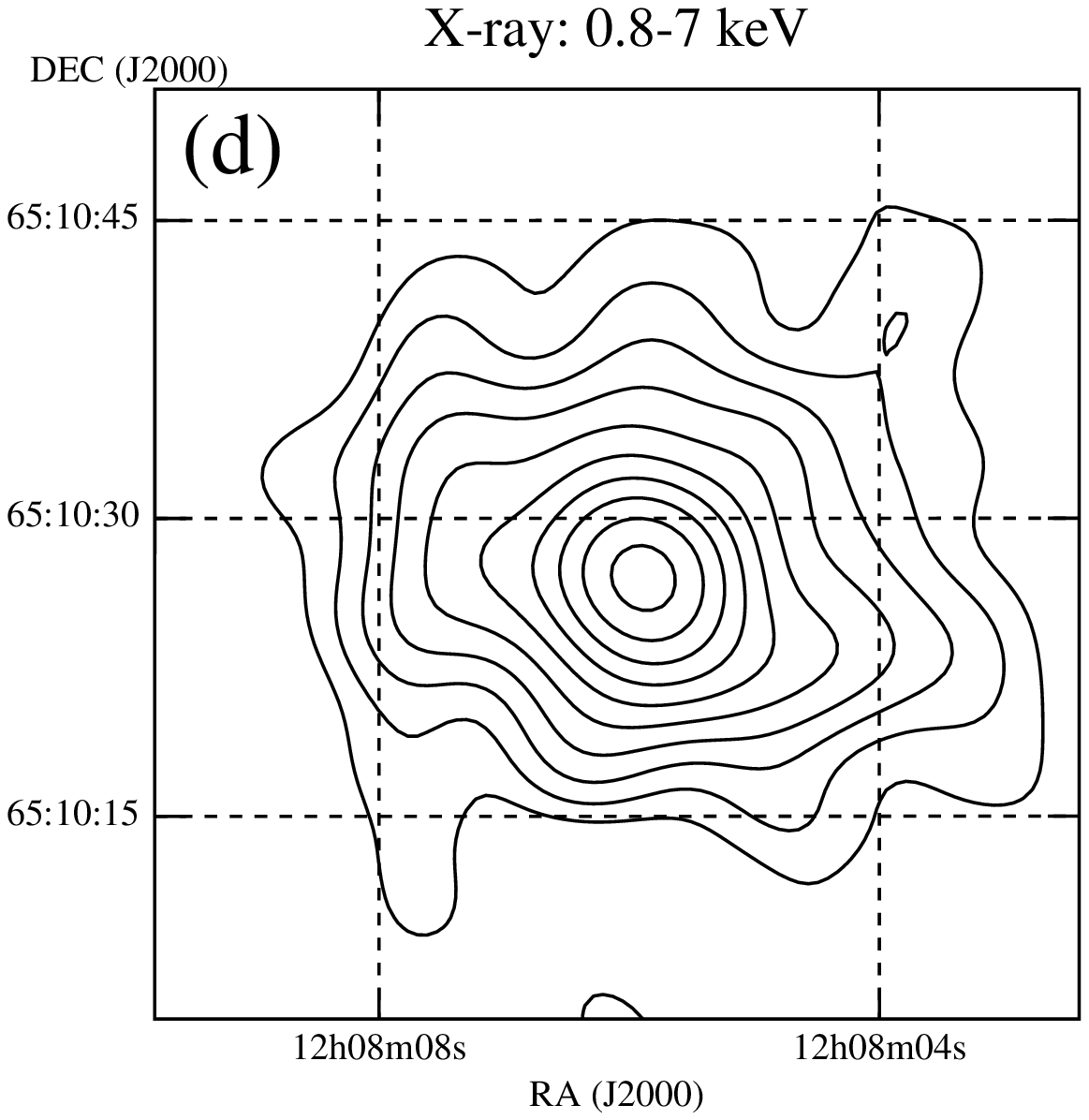}\\
\caption{Chandra X-ray smoothed image of NGC~4125. (a) The $0.3-7$ keV X-ray contour map after removing point sources and smoothing an image, overlaid on the original X-ray image. (b) The same contour map as in (a) on that of the $5.5-6.5$ $\mu$m continuum emission. (c) The $0.3-0.8$ keV and $0.8-7$ keV X-ray contour maps. The X-ray contours are drawn at logarithmically-spaced 10 levels from 80 \% to 3 \% for the panels (a) and (b) and from 80 \% to 10 \% for the panels (c) and (d).}
\end{figure}

\begin{figure}
\FigureFile(80mm,80mm){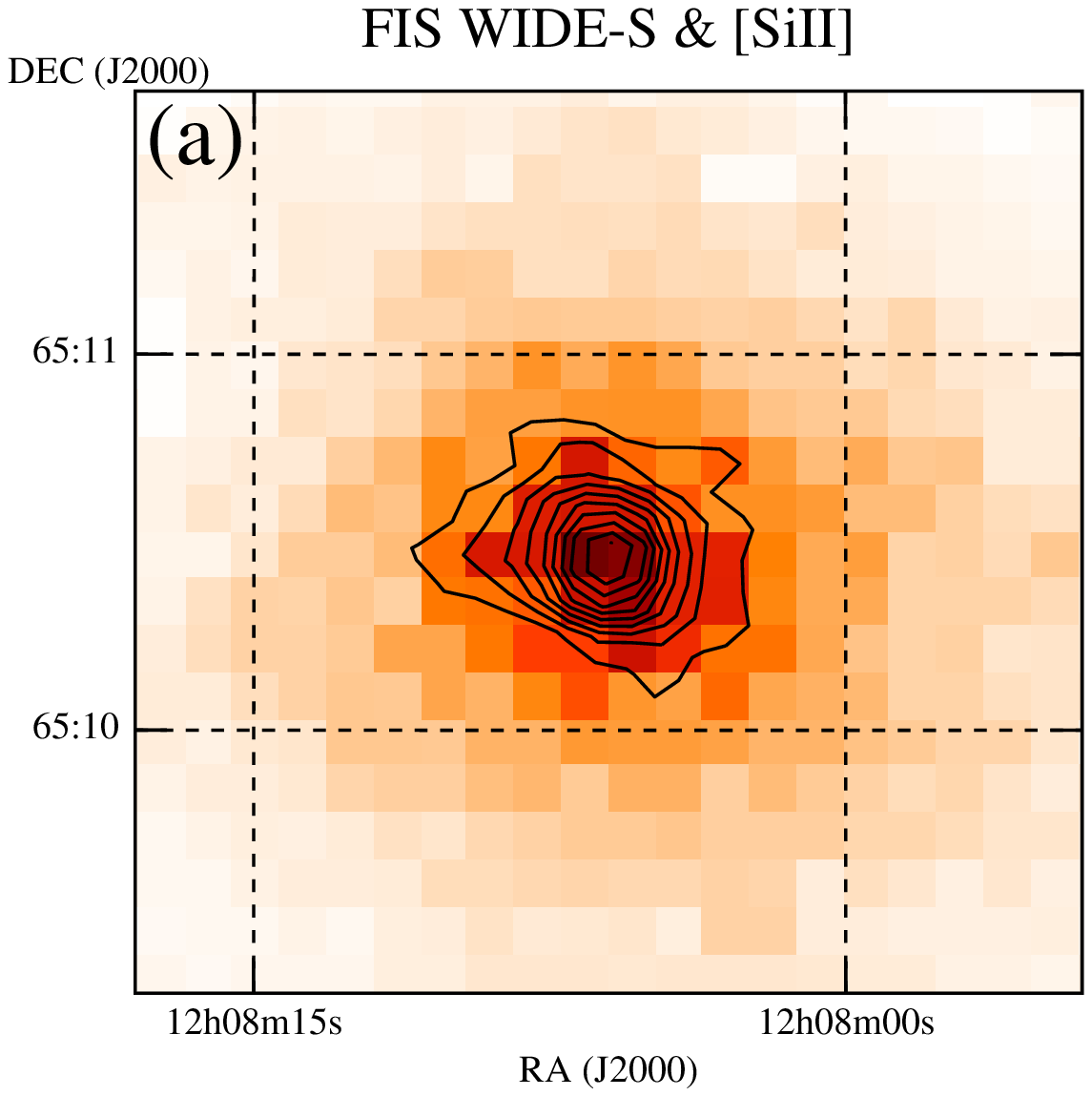}\FigureFile(80mm,80mm){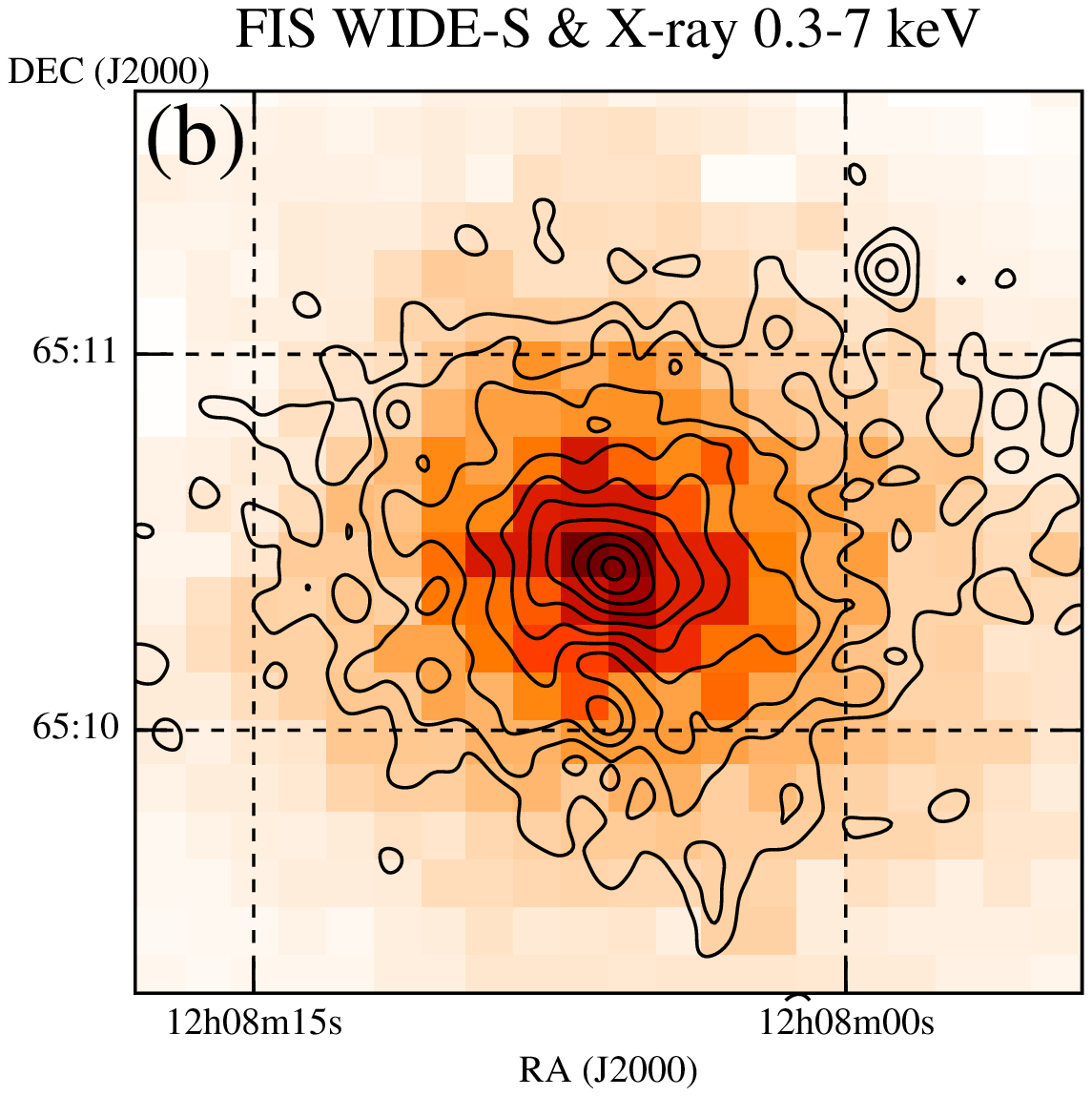}\\
\caption{(a) contour map of the [SiII] 34.8 $\mu$m emission, the same as in figure 2, superposed on the {\it WIDE-S} 90 $\mu$m image, the same as in figure 4, but displayed in the color levels. (b) The same X-ray contour map as in Fig. 8(a) superposed on the 90 $\mu$m image. Note that the image data of [SiII] are limited only to the central $45''\times45''$ area.}
\end{figure}

\subsection{Constraints on origins of dust and PAHs from their distributions}
Figure 10 summarizes the size and the axis direction of the spatial distribution of each spectral and photometric component in terms of the outer/inner flux ratio and the P.A. The size of the distribution is normalized to that for the 4 $\mu$m band intensity, i.e. the stellar emission. It is clear from the figure that the emissions by the PAHs as well as the molecular and atomic gases excluding the [SiII] emission are more centrally concentrated than the stellar emission, and on a larger scale, the emissions by the far-IR dust and the X-ray hot plasma are not more widely distributed than the stellar emission (probably less for the far-IR emission). All of these suggest that the PAHs and dust are not of an external origin; otherwise their distributions would be found to be extended more widely than the stellar distribution. In figure 10, we find that there are at least three types of axis directions. The continuum and the photometric band emissions dominated by the stellar emission possess the P.A. of $\timeform{82.5^{\circ}}$, while the atomic gas, dust, and hot plasma show $\sim 70^{\circ}$, a significanly smaller P.A. The PAHs and molecular gas show $\sim 90^{\circ}$, a significanly larger P.A. than the stellar emission.



\begin{figure}
\FigureFile(150mm,100mm){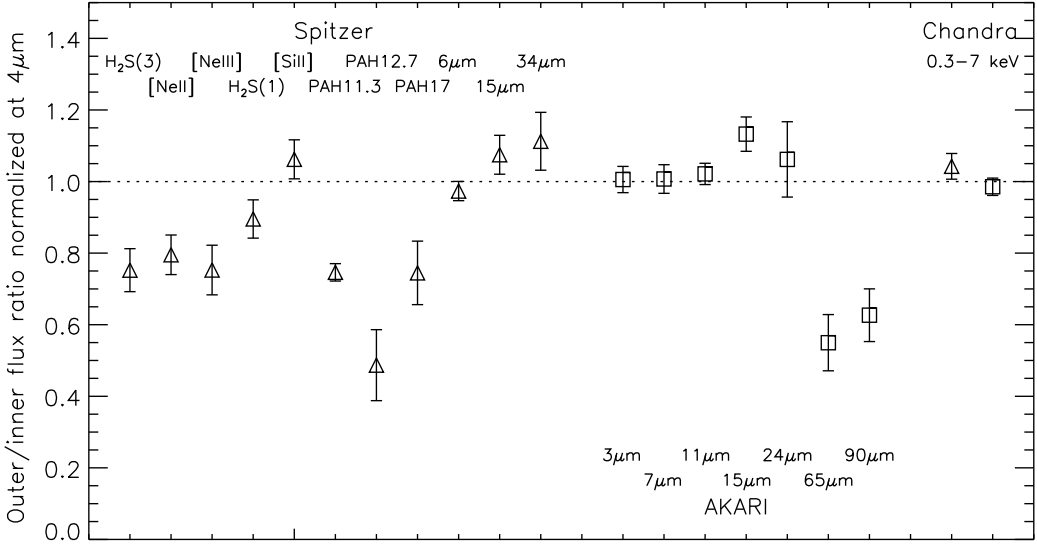}\\
\FigureFile(150mm,100mm){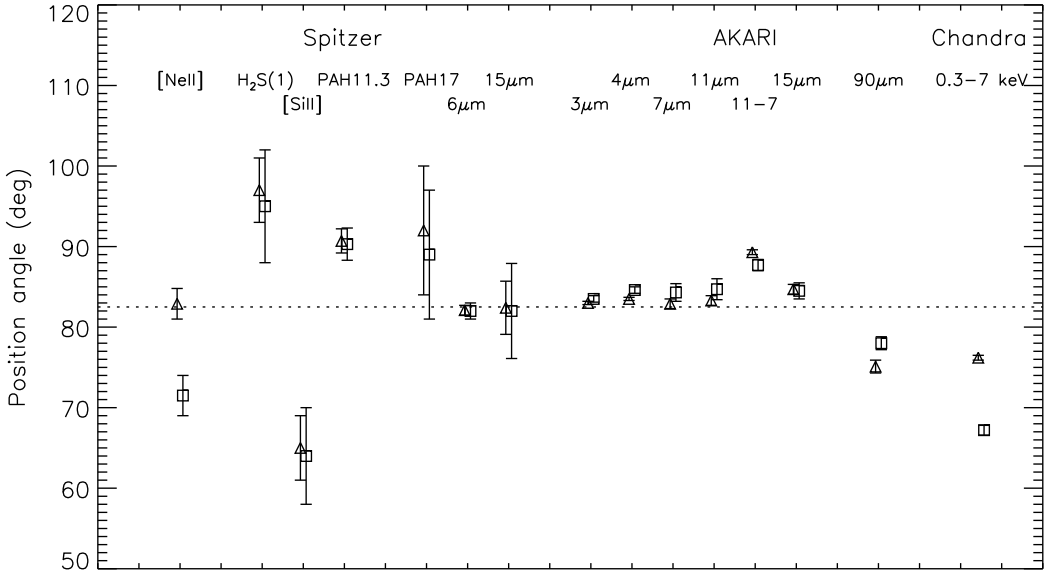}\\
\caption{(Top) Ratios of fluxes in outer to inner area for various spectral components and photometric band intensities, which are normalized to the corresponding ratios in the {\it N4} band. The triangles represent the ratios in the outer $25''\times25''$ to the inner $15''\times15''$ area, while the boxes are those in the $r=\timeform{1'.5}$ area to the inner $r=\timeform{14''.1}$ area. (Bottom) The position angles of the spatial distributions for the various components obtained by an ellipse fitting.  The triangles and the boxes represent the results of fitting to 15 \% and 20 \% brightness levels of the peaks, respectively. Only for the X-ray image are they 20 \% and 30 \% brightness levels. The errors are obtained by assuming the positional uncertainty of 0.5 pixels per bin. The dashed line corresponds to the stellar major axis of the galaxy (P.A.$=\timeform{82.5^{\circ}}$; Jarrett et al. 2003).}
\end{figure}

Near the center of the galaxy, the PAHs and dust are likely to have been protected from the interaction with the X-ray plasma inside the dense molecular gas; otherwise the PAHs would have been destroyed in a very short time scale of $\sim$100 yr (Micelotta et al. 2010). In other words, the hot plasma does not occupy the whole interstellar space near the center of the galaxy. In the outer region of the galaxy, however, from the similarity in their spatial distributions, the dust and hot plasma must be mixed to some extent, unless the volume filling factor of the hot plasma is small. It is very unlikely that the hot plasma has non-diffuse patchy distributions. In fact, our observation suggests that the dust is interacting with the hot plasma, producing cooler plasmas. As described above, dust grains with a 0.1 $\mu$m size can survive for a $10^6-10^7$ yr timescale, which is much longer than the lifetime of the PAHs but much shorter than the look-back time of $(4-6)\times10^9$ yr for the merger event (Schweizer \& Seitzer 1992). Therefore the dust is likely to have been supplied into the outer region of the galaxy quite recently apart from the merger event.

 
Our observation shows that PAHs are present only in the innermost region 
while the dust emission is obviously extended than the PAH emission. 
If we regard the observed PAHs as smallest form of dust grains, the observational result suggests the size-dependent distribution of dust, i.e. larger grains extended to outer areas, which favors an internal outflow origin for the dust-lane and extended materials. Temi et al. (2007) suggested that the buoyant outflow of dust from a central reservoir in the core of a galaxy may play an important role in supplying dust into the interstellar space of elliptical galaxies with a rather high buoyant velocity up to 400 km s$^{-1}$ possibly by receiving momentum from a jet emerging from a central nucleus. From this viewpoint, the extended dust and cooling gas emission structure might be related to the outflows from the LINER dusty nucleus of NGC~4125.    

 
\section{Conclusions}
We have performed mid-IR spectral mapping with Spitzer and near- to far-IR imaging observations with AKARI for NGC~4125, a nearby elliptical galaxy possessing diffuse interstellar X-ray hot plasma. We obtain the spatial distributions of dust and PAHs as well as molecular and atomic gases in NGC~4125. We also derive the distribution of diffuse hot plasma by removing point-like sources from Chandra X-ray images, and compare them to understand how they are separated or coexist in the interstellar space. We find that the spatial distributions of the PAHs and dust are different from each other. The PAH emission predominantly comes from a compact dense molecular gas region near the center of the galaxy, and hence the observed PAHs are likely to have been protected from the interaction with the X-ray plasma inside the dense molecular gas. There are no diffuse components observed in the PAH emission. In contrast, the dust and atomic gas emissions have more extended structures similar to the distribution of the X-ray plasma. The similarity suggests their association with the X-ray plasma. The PAH and the dust emission do not follow the stellar surface brightness, suggesting unimportance of stellar mass loss as their sources. NGC~4125 is thought to be a $(6-8)\times10^9$ yr old remnant of a major disk-disk merger. Even if dust and PAHs were brought in by the merger, they cannot survive until now considering their short destruction timescales in hot plasma. The PAHs and dust are likely to have been produced secondarily after the merger, having avoided their interaction with the hot plasma for a long time near the center of the galaxy. Recent transient events such as outflows from the central dusty nucleus may have spread the dust over the outer region of the galaxy.


\bigskip

We would express many thanks to an anonymous referee for giving us useful comments. This work is based on observations made with AKARI and Spitzer. AKARI is a JAXA project with the participation of ESA. Spitzer is operated by the Jet Propulsion Laboratory, California Institute of Technology under NASA contract 1407. We are grateful to the SINGS team for providing the CUBISM software and the PAHFIT tool. This work has also made use of data obtained from the Chandra Data Archive. This research is supported by the Grants-in-Aid for the scientific research No. 22340043, and the Nagoya University Global COE Program, ``Quest for Fundamental Principles in the Universe'', from the Ministry of Education, Culture, Sports, Science and Technology of Japan.








\end{document}